# The overlooked price feedback

*Why simulations overstate distribution grid peaks from home batteries under real-time pricing*

Lion Hirth

Hertie School, Berlin, and Neon Neue Energieökonomik, Berlin

hirth@hertie-school.org

**Abstract:** Simulation studies of home batteries under real-time pricing commonly find that synchronized charging creates new peaks in distribution grids. Yet all home-battery simulation studies reviewed here treat wholesale prices as exogenous, ignoring the effect of battery dispatch on prices. Charging in low-price hours raises the wholesale price, which dampens the incentive to charge. In equilibrium, charging is spread out across hours. A simulation of household battery dispatch under real-time pricing, calibrated to German data, finds that, if half of households own a battery, exogenous prices raise the coincident import peak by 83 % relative to the no-battery benchmark. With endogenous prices, the same fleet *reduces* the peak by 6 %. This finding is robust across alternative price functions, price paths, peak measures, and scenario assumptions. While new local peaks remain possible where batteries are geographically concentrated, these results suggest that published studies overstate the local peak-load effects of price-responsive batteries.

Keywords: real-time pricing; price feedback; endogenous prices; home batteries; distribution grid

JEL classification: Q41, Q48, L94, D47, C63

## Graphical abstract

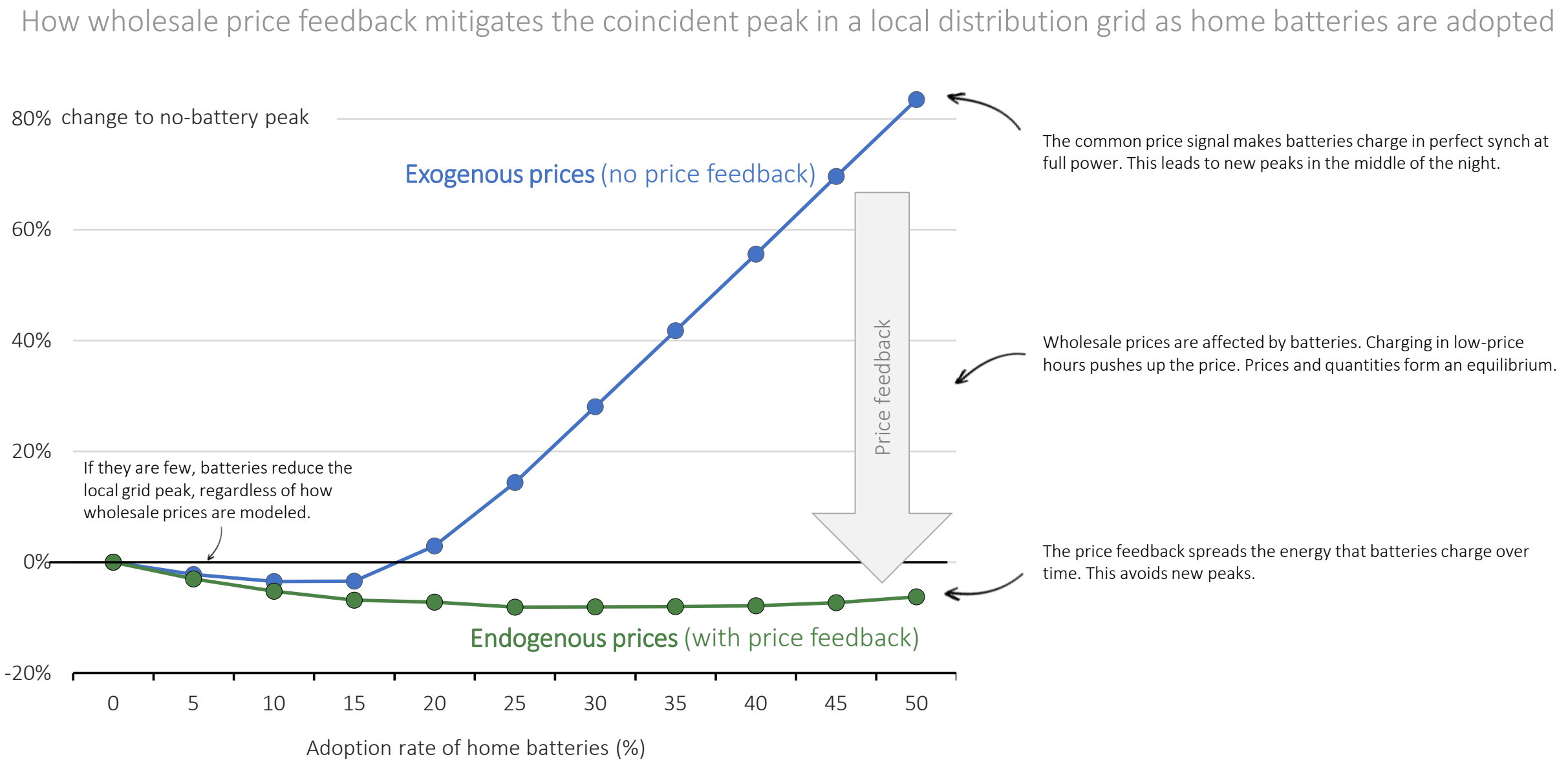

# 1. Introduction

Batteries are reshaping global power systems. This year, about 3 TWh of cell capacity will be produced, adding to the stock of around 7 TWh. The fastest-growing segments of battery deployment are electric vehicles and stationary energy storage. If this growth continues, batteries may provide most power system flexibility by mid-century, balancing electricity supply and demand across time scales from seconds to hours.

A growing number of households are installing stationary home batteries alongside rooftop solar panels. Reports suggest that about one in ten single-family homes in Italy, Austria and Australia owns a home battery. In some states of Germany, the share is likely to exceed 20 %. In the United States, home batteries are most widespread in Puerto Rico, where about 171,000 households and businesses had one at the end of 2025 (U.S. Energy Information Administration, 2026), and in Hawaii, where 92 % of residential solar permits on Oahu in 2025 included a battery (DBEDT, 2026); in California, more than 200,000 homes had a battery by late 2025 (California Energy Commission, 2025).

Home batteries offer their owners two financial benefits. First, they increase the share of rooftop solar generation consumed within the household, saving network charges and taxes. Second, under dynamic retail rates, they enable arbitrage: charging from the grid when electricity is cheap and potentially reinjecting it when prices are high. This is particularly relevant for retail rates with real-time pricing (RTP) that follow the hourly swings of wholesale electricity prices.

Distribution utilities (distribution system operators) are concerned that this behavior may increase stress on local distribution grids. Optimized home batteries and EV batteries under real-time pricing respond to the same national electricity price signal. This may lead to synchronized behavior, incentivizing all batteries to charge at full rated power in the same cheapest hour of the night (Muratori and Rizzoni, 2016). Because batteries have high power ratings compared to typical household peak demand (up to 11 kW in most markets), relatively few can create new peaks on distribution transformers. Almost all published studies find evidence for such “rebound” peaks. If that is true, policymakers face a trade-off between forgoing these batteries as a means of balancing the national electricity market and spending considerable sums on upgrading power grids.

Under real-time pricing, the wholesale electricity price determines battery dispatch. But the direction of causality runs both ways: battery charging adds electricity demand and affects the wholesale price. Low prices trigger charging, which increases demand and puts upward pressure on prices. The opposite is true for discharging. Prices and quantities are determined jointly in market equilibrium. This feedback spreads charging across hours. However, most published studies do not take this wholesale price feedback into account, potentially biasing their findings.

This paper assesses whether new local peaks from price-optimized battery dispatch persist when wholesale prices respond, or whether they are an artifact of assuming exogenous prices.

The paper makes three contributions. First, I systematically review the existing literature on local peaks triggered by dynamic pricing. Almost every study finds a new peak; not a single simulation treats prices as endogenous. Second, I quantify the effect of wholesale price feedback in household simulations. To endogenize prices, I specify them as a function of net load, including battery charging and discharging, and solve jointly for equilibrium prices and dispatch. Third, I derive a simple analytical model to identify the factors that determine if new local peaks occur under endogenous wholesale prices.

The simulations cover households with home batteries under real-time pricing and are calibrated to German data for 2025. At low adoption, batteries reduce the coincident peak both in the import and export direction. This is because batteries charge when prices are low, which tends to coincide with low local load or even local generation surplus. They discharge when prices are high, usually on high-load mornings and evenings. At very low power prices, all batteries simultaneously charge at full rated power; at high prices they all discharge. Wholesale prices fluctuate widely enough for such synchronized behavior to occur in hundreds of hours every year; the coincidence factor is virtually one. As soon as battery adoption reaches 15–20 % of households, synchronized charging creates new peaks during previously off-peak periods, e.g., winter nights. This “herding” or “overshoot” is in line with previous studies. The main finding of this paper is that these rebound peaks disappear once wholesale prices are modeled endogenously. The coincident peak remains below no-battery levels across all adoption rates, both in the import and export direction. The difference is economically significant: an increase in the coincident peak load of 83 % under exogenous prices becomes a *decrease* of 6 %.

I assess the robustness of these results using a range of alternative specifications of the price function, seven exogenous price curves, three alternative measures of the coincident peak, 22 scenario variations, three historical years, alternative temporal resolutions, and tests of the simulation procedure itself. In most specifications, the synchronized local peak disappears once wholesale prices are endogenous. One exception is local clustering: if batteries are concentrated in a small number of regions (e.g., affluent suburban communities) rather than spread out across many grids, peaks might emerge in those networks; the analytical model explains this finding. In this context, instruments such as curtailment by distribution utilities or congestion-based network charges might be necessary.

The next section reviews the literature. Section 3 explains price feedback in electricity markets. The methodology, simulation results and robustness checks follow. Section 7 discusses the results using an analytical model of new peaks under endogenous prices.

# 2. Literature review

From a pool of more than 360 papers, I identify 19 academic studies that quantify the impact of home batteries on the local electricity grid: 18 simulations and one randomized field experiment. Most were published in engineering, energy economics and interdisciplinary energy outlets. Table 1 provides details for each study. Table A1 lists comparable studies on other flexible loads such as electric vehicles. Table A2 lists studies reviewed but excluded from the comparison, e.g., because they do not provide a grid metric or do not optimize batteries according to prices. Appendix A2 describes the search and the selection criteria.

Table 1. Studies on home batteries

| Study | Setting | Price signal (retail rate) | Grid metric | Synchronized new peak? | Price feedback |
|---|---|---|---|---|---|
| Rodriguez-Garcia et al. (2026) | US; 178 synthetic feeders | TOU (seven US retail rate variants), with and without grid charging restrictions; net billing | Feeder peak (import); deferral value | Yes, on some feeders: investment rises on one feeder in five | No |
| Korab et al. (2026) | Poland; one low-voltage feeder, 144 customers | RTP: hourly Polish wholesale prices, identical for withdrawal and injection | Node voltage; overvoltage hours | Yes, export direction: overvoltage on a day with negative midday prices | No |
| Pimm et al. (2018) | UK; 150 aggregations of 100 households | TOU, three-tier; time-of-export rate, two-tier | Aggregate peak (import and export) | Yes, small: after-diversity maximum demand +2 % (without PV) | No |
| Stute and Kühnbach (2023) | Germany; suburban low-voltage grid, 146 nodes | RTP: EPEX day-ahead prices; or TOU, three-tier | Household peak (withdrawal); transformer loading; overload hours | Partly: overload hours up from 2 to 111 per year; grid maximum falls | No |
| Semmelmann et al. (2025) | Germany; 500 single-family households; ten medium- and low-voltage grids | RTP: day-ahead prices | Reinforcement cost (medium and low voltage) | Yes: reinforcement cost from 22 to 55 million EUR at full adoption | No |
| Cong et al. (2025) | Denmark; one 400 kVA feeder, 160 households | RTP: Nord Pool day-ahead prices, plus a TOU network charge | Transformer loading; overload hours (import) | Yes: transformer overloads within four months of the adoption path | No |
| Lilienkamp and Namockel (2025) | Germany; SimBench semi-urban grid, 12 low-voltage grids, 940 households | RTP: modeled quarter-hourly day-ahead prices; TOU derived from them | Transformer overload events (import) | Yes: about 4,800 overloads per year under TOU or RTP, none under the flat rate | No |
| Althaher et al. (2024) | Southern Jordan; real network, 2,440 residential customers | TOU: low price from 10 a.m. to 3 p.m. | Feeder peak (import); transformer loading; voltage | Yes: midday feeder peak of 16 MW against 9 MW originally | No |
| Lersch et al. (2025) | Germany; IEEE European low-voltage test feeder, 55 buildings | RTP: German day-ahead prices; variant with a time-varying network charge | Transformer and line loading (import); voltage; reinforcement cost | Yes: peak transformer loading from 29 to 70 % of rating | No |
| Riedel et al. (2024) | Southwest Germany; 74 single-family homes, 80 % with PV and battery | RTP: EPEX day-ahead price plus fixed surcharges | Aggregate peak (import) | Yes: aggregate peak from 105 to 263 kW | No |
| Brigger et al. (2026) | US; IEEE 123-node test feeder, 20 PV units with batteries | RTP: ISO-NE five-minute prices | Lost load under feeder thermal limits (import) | Yes: lost load rises above 10 % battery adoption | No |
| Young et al. (2019) | Australia (Sydney); 2,230 households | TOU; TOU with shoulder and demand tariff | Residential zone substation peak (import) | Partly: demand steps up when batteries stop at the end of the peak window | No |
| Fridgen et al. (2018) | US (Austin); 100 microgrids of ten homes | RTP: ERCOT prices; TOU; CPP | Peak at the microgrid connection point (import or export) | Yes: peak from 50 kW (flat) to 64 kW (RTP) with volumetric charges | No |
| Geske et al. (2026) | Luxembourg; 166 households on one transformer | RTP: hourly wholesale prices (TYNDP 2040 scenario), plus a capacity-subscription network tariff | Transformer overload hours (import and export) | Yes: 970 overload hours per year when 60 % of households respond | No |
| Restel and Say (2025) | Australia (Melbourne); 400 households with PV and battery | TOU import rate (peak 3 to 9 p.m.); time-varying export rate | Mean net load profile (import and export) | Yes: winter imports rise after 9 p.m.; exports surge at price transitions | No |
| Crowley et al. (2025) | Denmark; 15-node low-voltage community grid | RTP: spot prices | Import at the connection point | Yes: import peak between 4 and 6 a.m. above the capacity limit | No |
| Azuatalam et al. (2021) | Australia; three low-voltage feeders, 175 to 302 customers | TOU: network time-of-use tariff passed through to retail | Median customer monthly peak; feeder head loading; voltage (import and export) | Yes: off-peak charging raises the median monthly peak by up to about 10 % | No |
| Lopez Garcia et al. (2026) | Great Britain; randomized encouragement, 708 installations at low-income households | TOU: cheapest 2 to 5 a.m., peak 4 to 7 p.m. | Average household profile (import and export) | Yes: imports peak at 2 a.m., about 40 % above the pre-installation evening peak | No |
| Neumann and Blaschke (2026) | Germany; SimBench low-, medium- and high-voltage grids | Network charges only: critical peak pricing with event hours fixed ex ante; volumetric; peak demand charge | Annual peak; reinforcement cost (import) | Yes: under critical peak pricing the annual peak is a pre-charging hour | No: event hours fixed ex ante, independent of the response |

The studies measure the impact on the local grid in slightly different ways. The most common measure is annual coincident peak load at the feeder or distribution transformer, assessed in the import or export direction. Other studies assess voltage at the connection points or distribution grid reinforcement investments. This paper treats these measures as indicators of stress on the local distribution grid.

The review focuses on two questions: Do studies find a new peak, created by the synchronized behavior of flexibility assets? Do studies take into account the price feedback, i.e., are prices modeled endogenously?

Almost every study finds new peaks, except for very low adoption rates. Many find previous peak load doubles, some find it triples (Figure 1). None of the 18 simulation studies treats prices as endogenous; all of them ignore the wholesale price feedback.

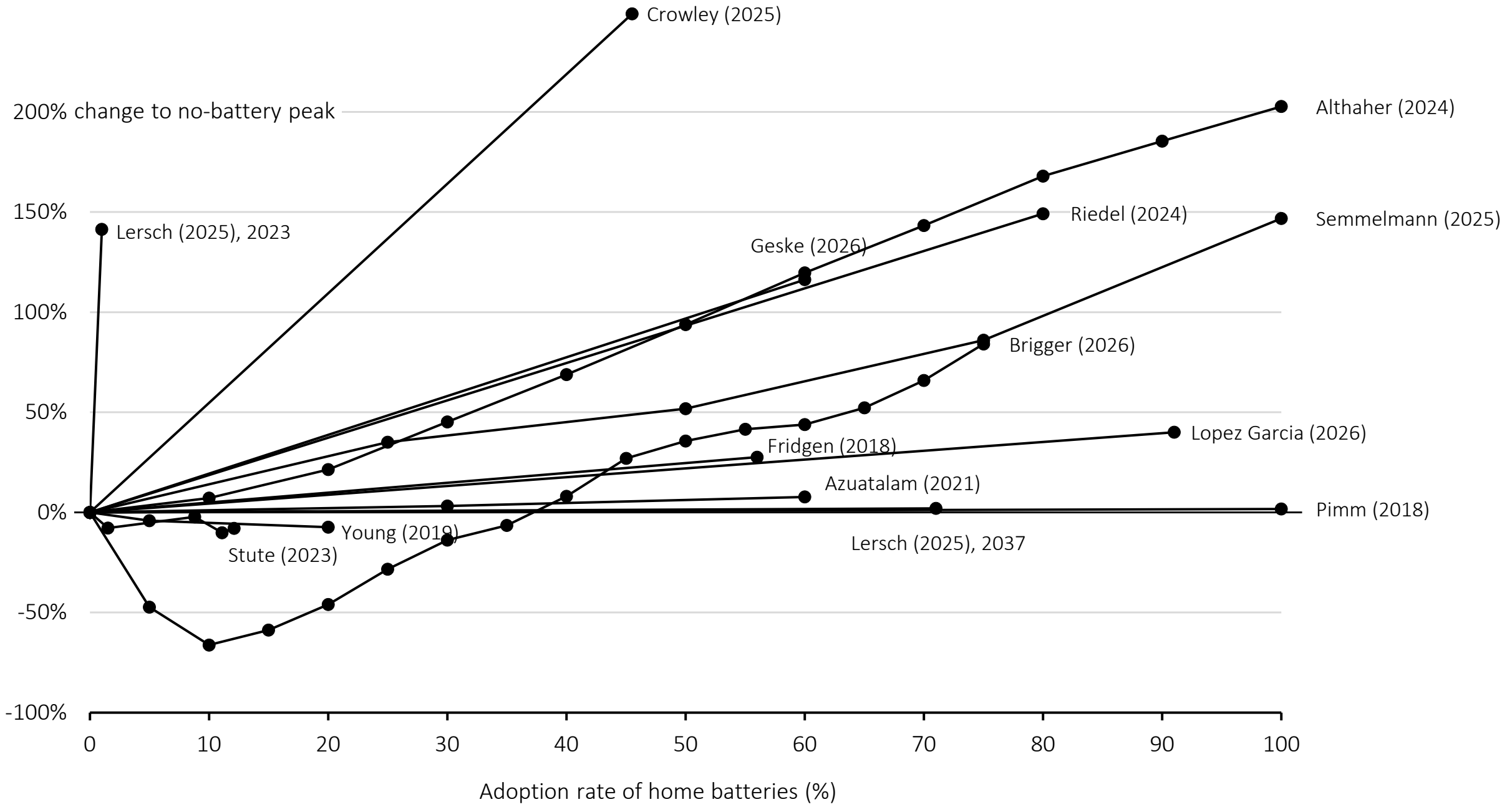

Figure 1. Local grid stress for increasing adoption of home batteries in previous studies.

# 3. Dynamic pricing and price feedback

Dynamic retail rates transmit wholesale price incentives to households through different contractual arrangements. This section explains how these arrangements accommodate price feedback and under what conditions equilibrium prices implement the corresponding battery dispatch.

## 3.1. Dynamic retail pricing

In liberalized electricity markets with unbundling, i.e., where grid operation is separated from retail supply and from power generation, the electricity bill includes three major components: the price of energy, network charge, and taxes and surcharges. Network charges are regulated because power grids are monopolies. The retailer sets the energy price based on its wholesale procurement costs. The contracting parties agree on the rate, including whether the price is constant or varies during the day.

This paper focuses on the energy component of retail rates because the price feedback studied here operates through wholesale markets, not network charges.

Under a dynamic retail rate, the price varies over the contract period, usually within the day. This includes four major types of contracts:

- Time-of-use pricing – predetermined, fixed periods with prespecified prices such as peak and off-peak prices.
- Real-time pricing (RTP) or “spot tariffs”, where the retail rate includes the hourly or subhourly wholesale price, usually the day-ahead price, plus a markup.
- Critical peak pricing, where the price is generally constant except in rare, preannounced periods of scarcity when it is much higher.
- Flexibility contracts, where utilities get the right to optimize an asset in return for a discount, e.g., the right to trade an EV battery on the market for eight hours per day in return for free charging of the car (Table A3).

All these rates may be combined with a constant volumetric network charge, a time-varying one, or none.

With the rise of rooftop solar photovoltaics, electric vehicles and home batteries, households are increasingly gaining the technical capability to optimize withdrawal of electricity from and injection into the grid – through load shifting, storage and, at times of negative wholesale prices, curtailment of solar generation. The economic purpose of dynamic pricing is to use this flexibility to balance electricity supply and demand efficiently over time (Borenstein, 2005; Joskow and Wolfram, 2012).

## 3.2. Wholesale price feedback

The mechanism at the heart of this paper is the feedback from battery dispatch to the wholesale electricity price. A fleet of batteries that charges in the cheapest hours adds demand precisely in those hours. This raises the price to which the fleet responds. Electricity demand and price are therefore jointly determined, endogenous variables. Modeling them requires solving a system of equations simultaneously rather than assuming that price determines dispatch without feedback. Prices and quantities form an equilibrium that must simultaneously fulfill two conditions:

- Wholesale prices are consistent with overall demand and supply, including battery (dis)charging
- Battery dispatch is optimal given these wholesale prices, i.e., batteries have no incentive to deviate

Under flexibility contracts, where utilities optimize assets directly, the feedback is direct: buying energy on the wholesale market to charge a battery raises the price; selling discharged energy lowers it.

For most dynamic retail rates, however, including time-of-use and real-time pricing, the sequence of events initially seems incompatible with wholesale price feedback. The price is set in advance, whether the day before or a year earlier, and remains fixed afterward. Only then do households choose their battery dispatch. How can that dispatch feed back to wholesale prices?

The answer is expectations. Retailers anticipate their customers’ response and incorporate it into their wholesale bids. They have a direct financial incentive to do so: a retailer that ignores the battery response buys too little energy in low-price hours, when the batteries charge, and pays for the shortfall at imbalance prices. In practical terms, instead of buying a fixed quantity through unlimited bids in day-ahead auctions, they may submit price-sensitive bids to procure more energy at lower prices, anticipating higher consumption. Alternatively, or in addition, they may engage in short-term electricity procurement on intraday, real-time, and imbalance markets after day-ahead prices are known to them and their customers.

The retailer need not form accurate expectations itself. If it ignores the price response of home batteries, the quantities will show up at later market stages, i.e., the intraday market or imbalance settlement. Trading houses can detect such patterns and take positions on the day-ahead market to profit from these price spreads. As a result, the expected battery dispatch enters the day-ahead market through the bids of those trading houses.

The model used in this paper (see Section 4.4) does not distinguish between these channels: the price function is a reduced form of the resulting expectations equilibrium. It assumes that retailers anticipate the aggregate battery response correctly, so that day-ahead prices already reflect equilibrium.

## 3.3. On the existence of a market equilibrium

Under real-time pricing, prices and dispatch are determined in sequence. First, the day-ahead auction sets wholesale prices based on bids of retailers that anticipate the response of their customers. Second, the retail price, i.e., the wholesale price plus network charges, is set and does not change afterward. Third, households optimize battery dispatch against the retail price. In other words, the retail price is a posted price: it is fixed before dispatch and does not respond to the quantities households choose. A time-varying network charge is a posted price in the same sense. In a day-ahead auction, the retailer bids the expected battery response of its customers.

If the price function is strictly increasing in each hour and the battery problem is convex, a wholesale market equilibrium exists and the clearing prices are unique. Whether prices also uniquely determine dispatch depends on the shape of the battery response. With heterogeneous batteries, strictly convex cycling cost, or a fleet small enough to charge at full power in the cheapest hours, the response is single-valued and prices determine dispatch. Under a posted price, this is the condition for an equilibrium: the retailer can anticipate the response exactly because it is unique.

With identical batteries and linear cycling costs, prices do not uniquely determine dispatch. A large fleet of such batteries equalizes prices across the hours in which it charges below full power; the aggregate response to the price of these hours is horizontal. At equal prices households are indifferent about when to charge, so prices do not determine the allocation of charging across these hours. Under a posted price, an equilibrium then exists only in a weak sense: the bid allocation is one of many optimal responses, but nothing in the price makes households choose it.

If one hour is marginally cheaper, all batteries move there. The posted price does not respond to this shift, so the bid volumes are wrong. Once retailers anticipate the shift, that hour becomes the most expensive one. Dispatch then flips back and forth between hours instead of converging to the fixed point (oscillation in Heid et al., 2026; herding under open-loop prices in Flath and Gottwalt, 2016; iterative price coordination in Gan et al., 2013 and Ma et al., 2013).

These considerations matter for real-time pricing, not for flexibility contracts. Under a flexibility contract, the retailer dispatches the batteries itself. There is no posted price and no third step: the retailer can implement the allocation it bid, including one of many allocations at equal prices. The simulations in Section 4 compute the fixed point directly, i.e., they assume that the retailer bids it and that households implement it.

# 4. Methodology

This paper simulates how optimized home batteries behave under real-time pricing. This section introduces the household-level simulations, the measure of the coincident peak, and wholesale price modeling.

## 4.1. Household properties

I simulate the electricity consumption, generation and storage of 100 households, a typical size of a German low-voltage network, at hourly resolution for one year to calculate the coincident peak for imports and exports (Section 6.3 reports robustness with respect to household population and bootstrap resampling). If these households are located on the same local distribution grid, this corresponds to the loading of the distribution transformer. Half the households own solar panels, a varying share of which have home batteries. All of them are assumed to be on real-time pricing, i.e., they pay the hourly wholesale day-ahead price plus a fixed volumetric network charge.

The 100 households are single-family homes with synthetic load profiles based on 2025 weather with daily shapes resembling the German standard load profile for residential consumers (Fünfgeld and Tiedemann, 2000), assuming 20 % adoption of heat pumps and EVs with uncontrolled charging. This results in a coincident peak of 169 kW that occurs on a February evening at 7 p.m.; the 1 % of hours with the highest load all fall on winter evenings. Figure A1 shows individual household load and aggregate load during the week of the annual import peak. In sensitivity runs (Section 6.3), different synthetic and empirically measured profiles are tested.

Half of the households own a PV system of 8 kWp. Generation is based on ERA5 reanalysis irradiance for 2025 near Berlin, assuming four orientations (east, south, west and flat). For generation injected into the grid, households receive the wholesale price. PV generation is not curtailed, even at negative wholesale prices. The export peak of 208 kW occurs on 20 June at 1 p.m. Sensitivity runs test different PV specifications (sizes, orientation, curtailment). Table 2 summarizes the resulting household characteristics.

Table 2. Household characteristics

| Statistic | Households | Min | Mean | Max |
|---|---|---|---|---|
| Annual consumption, kWh | All 100 | 1,378 | 5,428 | 15,331 |
| Annual PV generation, kWh | 50 with PV | 6,184 | 7,111 | 8,458 |
| Highest hourly withdrawal, kW | All 100 | 3.5 | 7.2 | 17.7 |
| Highest hourly injection, kW | 50 with PV | 5.0 | 5.3 | 6.0 |

Only solar households can own batteries, which are assumed to have an energy capacity of 8 kWh, rated power of 4 kW, a round-trip efficiency of 90 % and a cycling cost of 20 EUR/MWh. Battery adoption is varied between zero and 50 %. For robustness, I test different battery sizes and cycling costs and allow households to own batteries without solar.

## 4.2. Household optimization problem

Each household solves an optimization problem under perfect foresight (equations in Appendix A4); for numerical reasons, each month is optimized separately. Given the hourly retail rate, it minimizes the annual electricity bill. For each hour, the household chooses how to operate its battery. Prices are as follows:

- The retail rate is the wholesale price plus a fixed payment of 0.10 EUR/kWh for network charges (and taxes and surcharges) which applies to withdrawals (varied in sensitivities).
- When injecting solar energy into the grid, the household earns the wholesale price but neither pays nor receives a network charge.
- The battery may charge from the grid and inject into it. For electricity that is charged from the grid and reinjected, the household earns the wholesale price plus the refunded network charge; losses are not refunded.

In this setting, batteries reduce electricity bills through two mechanisms, self-consumption and arbitrage:

- Self-consumption of behind-the-meter solar generation avoids the volumetric network charge applied to electricity drawn from the grid. In my simulations, this “tax gap” between wholesale and retail prices is 0.10 EUR/kWh, making up slightly more than half the electricity bill. Adding a battery allows a larger share of that generation to be consumed within the household. The larger the tax gap, the more attractive self-consumption becomes for the individual household.
- By charging the battery at low prices and discharging it at higher prices, the household earns money from arbitrage. Because network charges are reimbursed, losses and cycling costs are the only barriers to arbitrage. Arbitrage becomes more attractive with wholesale price spreads.

The models use hourly resolution because wholesale prices were hourly until September 2025. A sensitivity run uses quarter-hourly granularity.

## 4.3. Measures of grid impact

Published studies use a wide range of metrics to measure the impact of household behavior on local distribution grids, e.g., voltage at the connection points, incidents of overload, or distribution grid reinforcement investments (recall Table 1). The most common measure is annual coincident peak load. This is what this paper uses, too.

More precisely, my preferred measure of grid impact is the mean aggregated load (kW) in the highest 1 % of hours, separately for import and export. I prefer this measure to the single highest hourly load because the extreme upper tail may not be representative. Transformers also tolerate short overloads. In sensitivity runs, the absolute peak (highest hour), the top 10 % and the 4-hour peak are also reported.

## 4.4. Market modeling and price feedback

Battery dispatch and wholesale prices are determined jointly. Wholesale prices $p_t$ are a function of net load after flexibility (including home storage) $R_t$ plus an otherwise unexplained residual $\varepsilon_t$:

$$p_t = f(R_t) + \varepsilon_t \tag{1}$$

Net load after flexibility is national load $L_t$ (including commercial and industrial demand, pumped-storage losses and the daily mean of net exports) minus wind $W_t$ and solar $S_t$ generation minus battery dispatch $b_t$:

$$R_t = L_t - W_t - S_t - b_t \tag{2}$$

Battery dispatch is the sum of discharging $d$ minus charging $c$ of both utility-scale and home batteries (superscripts $u$ and $h$) across $N$ identical local distribution grids (190,000 in the base case):

$$b_t = (d_t^u - c_t^u) + N \cdot \sum_{i=1}^{H} \left(d_{i,t}^h - c_{i,t}^h\right) \tag{3}$$

$R_t$ is also the load that thermal power plants have to serve, i.e., the thermal generation. The wholesale price is the marginal cost of the last plant needed to serve it, so the price rises with $R_t$ along the supply curve of the thermal fleet. For the simulations, a parameterized price function is needed. My preferred specification is a cubic price function:

$$f(R_t) = a \cdot {R_t}^3 \tag{4}$$

where $a$ is a parameter calibrated to historical prices. For robustness, I use two additional functions. The second specification is a quadratic function:

$$f(R_t) = a_2 \cdot \operatorname{sgn}(R_t)\,|R_t|^2 \tag{5}$$

The third specification is a piecewise linear function with ten segments and breakpoints at the deciles of thermal net load.

The parameters $a$ and $a_2$ of the cubic and the quadratic function and the slopes of the piecewise linear function are estimated by regressing hourly German 2025 day-ahead prices on thermal net load with day fixed effects (all data from ENTSO-E). The piecewise linear function is continuous with non-negative slopes. The residual between the observed price and the fitted price function is added to the price as an exogenous hourly price component. To represent how utility-scale batteries, pumped hydro, imports, exports and other flexible resources respond to changing price patterns, I model them collectively as utility-scale storage with 10 GW of power and 60 GWh of energy capacity and a cycling cost of 40 EUR/MWh. The properties of the prices resulting from this procedure are reported in Table 3.

Table 3. Properties of the observed 2025 day-ahead price and of the three price functions at zero home storage adoption

| Property | Observed 2025 | Cubic | Quadratic | Piecewise linear |
|---|---|---|---|---|
| Calibration parameter | n.a. | $1.8 \times 10^{-3}$ | $8.9 \times 10^{-2}$ | (many) |
| Mean, EUR/MWh | 89 | 91 | 91 | 90 |
| Minimum, EUR/MWh | -250 | -254 | -258 | -257 |
| Maximum, EUR/MWh | 583 | 533 | 549 | 549 |
| Mean daily spread, EUR/MWh | 124 | 119 | 121 | 122 |
| Hours with negative price | 576 | 617 | 635 | 686 |

# 5. Results

This section compares battery dispatch and coincident peaks under exogenous and endogenous prices. It then examines different battery configurations, local concentrations of batteries, renewable scenarios, network charges and EVs in place of stationary home batteries.

## 5.1. Battery dispatch

The wholesale price feedback changes battery dispatch considerably. Figure 2 (top) shows the dispatch of the fleet of home batteries for exogenous prices (blue) and endogenous equilibrium prices (green) for five days compared to the load of the local grid. With exogenous wholesale prices, battery dispatch is spikier. Occasionally, the batteries charge and discharge in perfect sync, i.e., all in the same direction at full power. When prices are allowed to respond to battery dispatch, they dampen those incentives. Charging and discharging are smoother. Full-power synchronization becomes much less frequent.

This can also be seen in the national net load after flexibility $R_t$ (bottom). With exogenous prices, batteries regularly trigger new peaks in national net load; with endogenous prices, the net load curve is flatter, not only compared to the exogenous case but also compared to the no-battery benchmark.

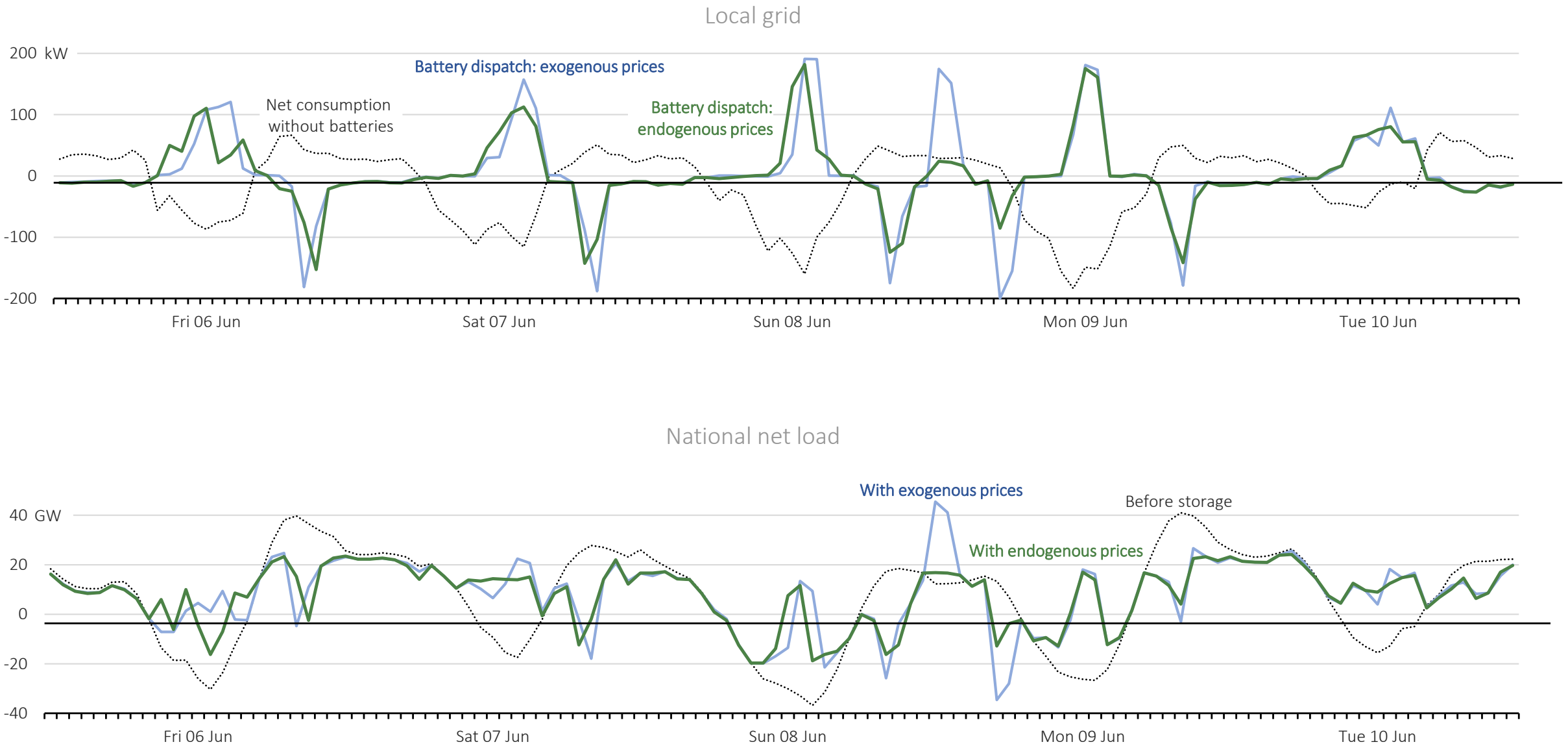

Figure 2. Battery dispatch and resulting national net load under exogenous and endogenous prices during five days.

This is also visible in the duration curve of battery dispatch, i.e., the sorted charging of the battery fleet across the year (Figure 3). Without price feedback, the battery fleet charges and discharges in perfect sync in 169 hours. With price feedback, this number is reduced to 7 hours. These remaining hours of synchronized dispatch tend to reduce the local coincident peak, for example through charging during solar peaks.

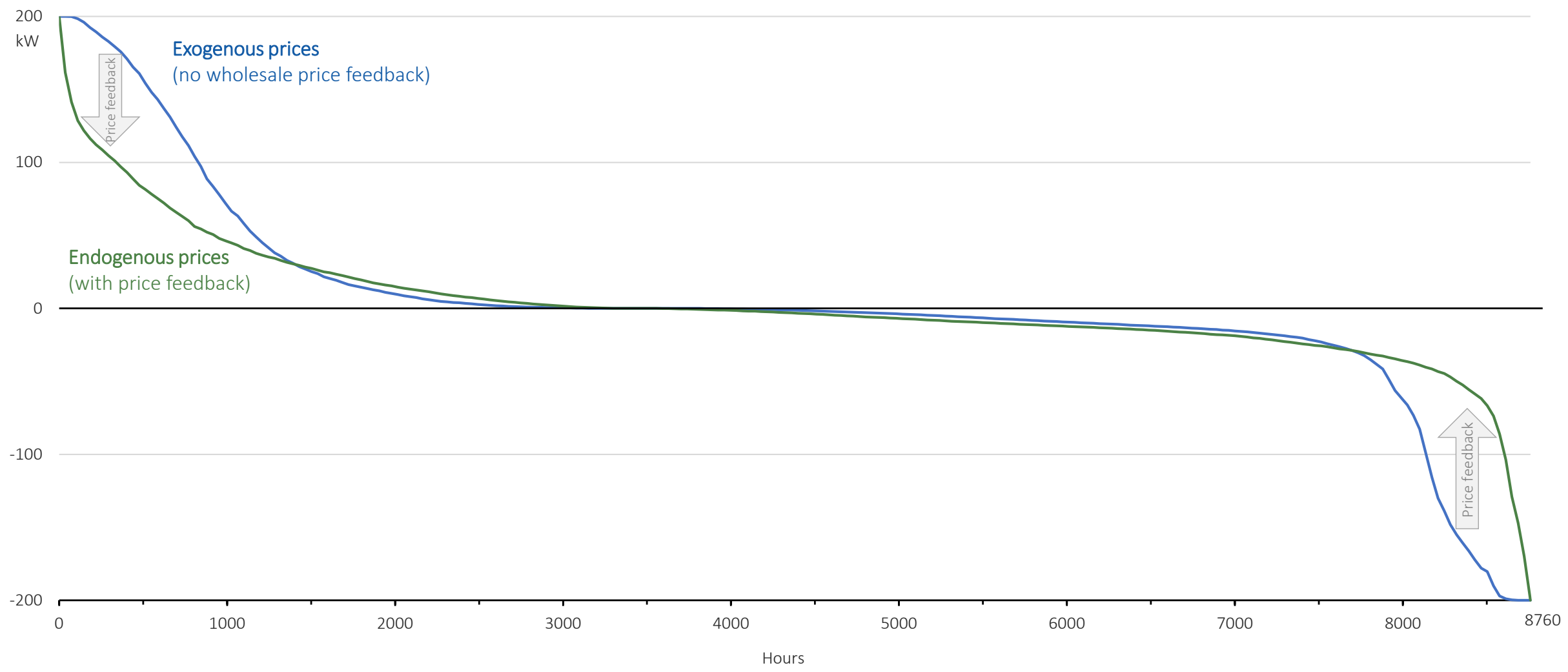

Figure 3. Dispatch of the local battery fleet, ordered by size (duration curves).

There are two mechanisms through which the price feedback reduces the coincident peak of batteries. First, even in the hours when they charge in sync, they do so at lower power. As the fleet charges into the hours of low net load, it raises the price in those hours until the price is flat across the charging window.

Second, the price feedback also reduces the synchronization across individual batteries. At full adoption, the mean pairwise correlation of hourly dispatch between two batteries is 0.88 under exogenous prices and 0.54 under endogenous prices. Under exogenous prices, wide price spreads make arbitrage between charging from and reinjecting into the grid profitable for virtually every battery. Under endogenous prices, spreads are compressed such that arbitrage is profitable less often: the share of charging for reinjection falls from 44 % to 20 %. Instead, self-consumption dominates the battery dispatch decision, which depends on the individual load and PV profile of each household. Because these profiles are heterogeneous, individual batteries charge and discharge at different times.

## 5.2. Main result: coincident peak

Figure 4 summarizes the main result of this paper. It shows the annual coincident peak (top 1 % of hours) for growing adoption of home batteries. At low adoption rates, batteries reduce the peak, regardless of how they are modeled: they tend to charge when local demand is low or negative, because this tends to coincide with low or negative national wholesale prices. In turn, they discharge at high prices, which correlate with local demand. Batteries thus reduce both the import peak (left) and the export peak (right) under exogenous prices (blue line) and endogenous prices (green line). As long as the number of batteries is small, whether prices are modeled endogenously or exogenously does not matter, because total battery capacity is too small to shift prices.

This changes as battery adoption grows. If prices are assumed to be exogenous, the synchronous pattern of high-power charging and discharging creates new peaks as soon as an adoption threshold is crossed. The threshold is 15–20 % for imports and 45–50 % for exports. The new peaks occur in what used to be off-peak periods, usually on winter nights (Figure A2 in the Appendix). This, however, is only true as long as wholesale prices are not allowed to change. Endogenizing wholesale prices results in import coincident peaks stabilizing well below the no-battery case and export peaks continuously falling with more batteries added. The difference is large. At full battery adoption, the price feedback reduces the import peak by about half and the export peak by 15 % (Table A5 in the Appendix).

The simulations thus reproduce the new peaks found in almost every study that assumes exogenous wholesale prices. They also suggest that these findings may be explained by the omission of wholesale price feedback.

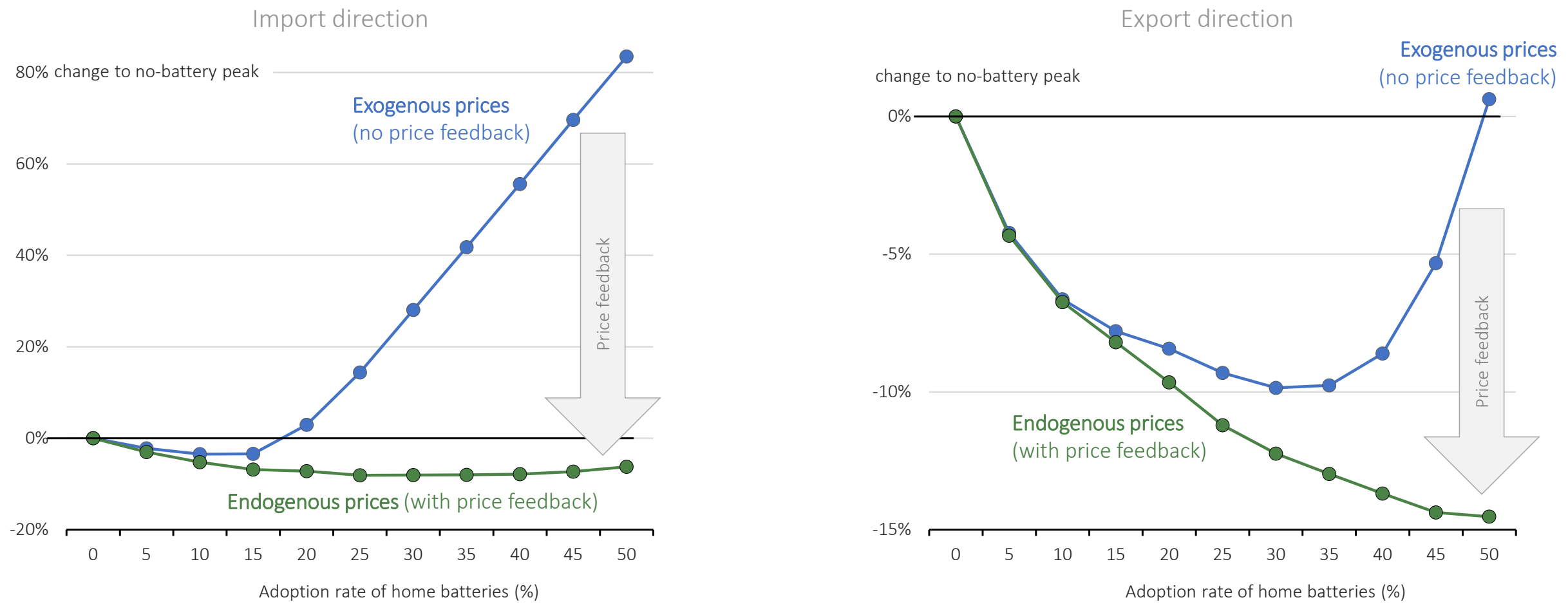

Figure 4. Coincident peak for exogenous and endogenous prices, import and export direction.

## 5.3. Battery size and adoption

This section presents sensitivity runs in which battery properties are varied: battery size is halved and doubled (larger batteries are common in other markets; in Australia, batteries installed under the federal rebate program average about 25 kWh; Energy-Storage.news, 2026); battery charging power is halved and doubled; cycling costs are increased; and batteries are also adopted by households without solar panels (see Table A5 in the Appendix for parameters). Figure 5 shows, for each variation, the coincident peak for exogenous prices (blue) and endogenous prices (green); Table A5 in the Appendix reports the values at full adoption.

In every variation, the price feedback reduces the coincident peak both in the import and the export direction. As expected, smaller batteries, reduced charging power and higher cycling costs imply that the new peaks that batteries create under exogenous prices are smaller; the peak-reducing effect of price endogeneity is diminished accordingly (but still present). Larger batteries, higher power ratings, and greater battery adoption lead to more extreme new peaks; accounting for price feedback still reduces them, but does not always push peaks below the no-battery level. Price feedback is therefore no guarantee against new peaks. These results are consistent with the analytical model of Section 7: under endogenous prices, the winter-night charging peak depends on battery capacity, not on rated power.

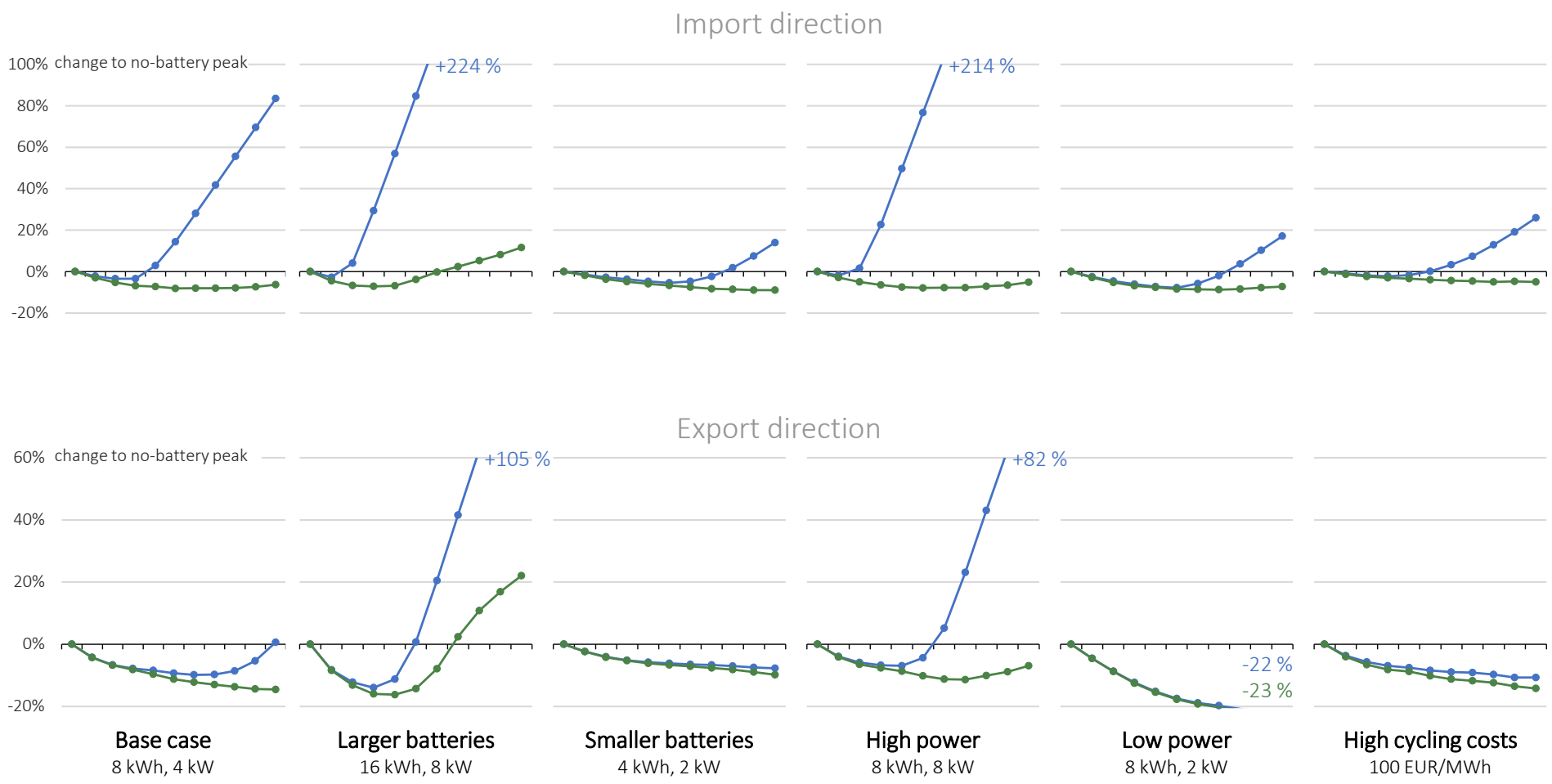

Figure 5. Coincident peak for exogenous and endogenous prices: battery size and adoption.

## 5.4. Local concentration

The impact on distribution grid assets is a local phenomenon, while the price feedback happens at the level of the wholesale market. In Germany, with its uniform wholesale market, that is the entire country; in other jurisdictions it would be bidding zones or, under locational marginal pricing, transmission nodes. In any case, the wholesale market covers a much larger area than a local distribution grid. As long as local grids are similar in load patterns and battery adoption, the wholesale price feedback smooths local peaks nevertheless.

The positive grid impact of real-time pricing, however, rests on the assumption that the flexible assets are relatively uniformly distributed. Only then does the wholesale price feedback work well in smoothing charging and discharging, avoiding new peaks. If battery adoption concentrates in local clusters, e.g., affluent suburban neighborhoods dominated by single-family homes, this is no longer the case.

In the simulation, this is adjusted by setting $N$ in Eq. (3), the number of identical grids. In the base case, it is set to $N = 190{,}000$. This number derives from the housing stock, assuming that home batteries are installed in specific types of buildings: 13.5 million dwellings in single-family and 5.5 million in two-family houses in Germany at the end of 2024 (Statistisches Bundesamt, 2025), divided by 100 households per grid. At full adoption, half of all single- and two-family homes in Germany therefore have rooftop solar and a battery. These batteries are distributed evenly across the country, with total power of 38 GW and energy capacity of 76 GWh.

To test for the effect of a more uneven allocation of batteries, $N$ is varied between 25 % and 200 % of the base case (Table A5). As expected, a higher concentration of batteries in fewer clusters leads to new peaks in those areas (Figure 6). This is consistent with the analytical model of Section 7: halving the number of grids doubles the share of each grid and halves national flexibility, which raises the local charging peak by a factor of √2 (Equation (9)). Modeling home batteries in a single network $N = 1$ replicates the exogenous price case. A similar but less pronounced effect occurs if utility-scale flexibility is scaled (Figure A3 in the Appendix).

That utility-scale flexibility matters so little is a result of the cost assumptions. Home batteries pay cycling costs only on discharge and, thanks to the refund, a network charge only on their losses; utility-scale storage pays cycling costs in both directions. Home batteries therefore accept a higher nighttime price than utility-scale storage. Once their charging lifts the nighttime price plateau above the level at which utility-scale storage still charges, they crowd it out of the low-price hours, so that the local charging peak depends only on the size of the home battery fleet and the shape of national net load. Utility-scale storage then affects home battery dispatch only through its discharge, which depresses evening prices. With the opposite cost ordering, utility-scale storage would occupy the lowest-price hours first and home batteries would charge less in those hours.

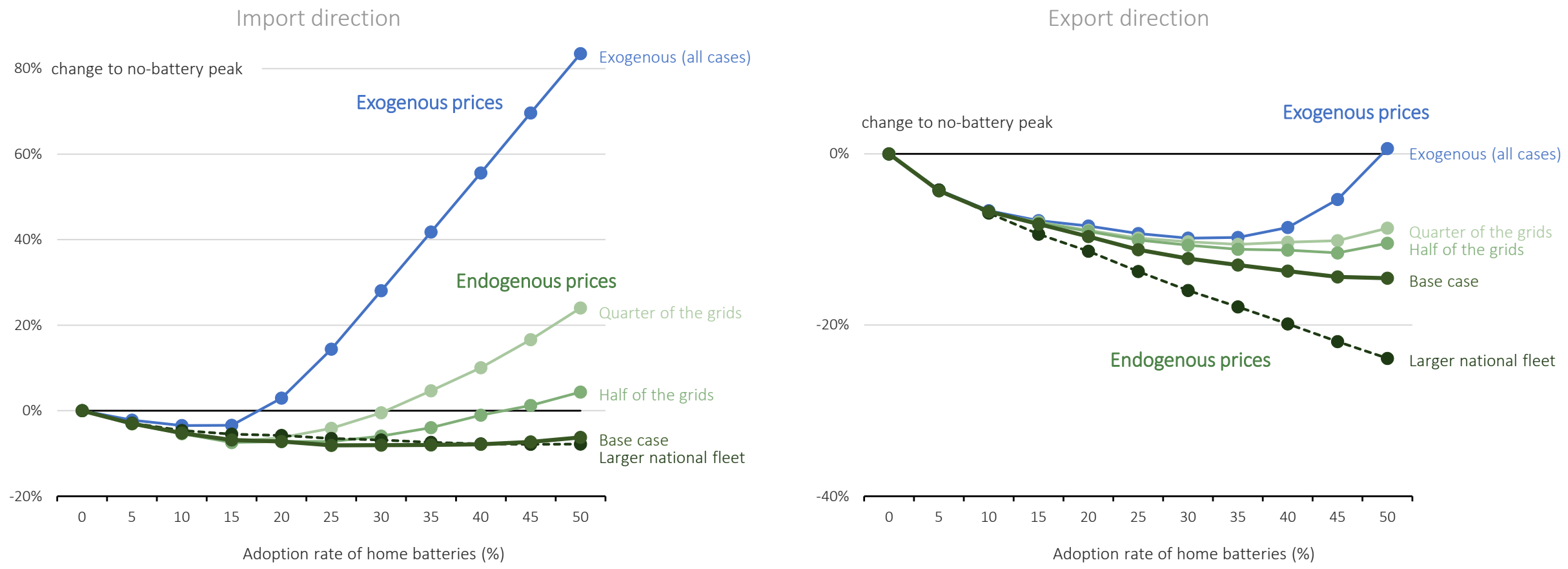

Figure 6. Coincident peak for exogenous and endogenous prices: local concentration.

## 5.5. Renewables

Wind and solar generation shape net load and hence wholesale prices. Residential solar also impacts the peaks in the local grid and the self-consumption incentives of households. To test for those factors, I allow for economic PV curtailment, double the size of residential solar panels (which doubles annual PV generation per household and also approximates sunnier climates), increase utility-scale wind and solar supply, and model a "2035" scenario with electrification and more renewables across the board (Table A5).

None of these assumptions materially alters the coincident peaks and the difference that endogenous prices make (Figure 7).

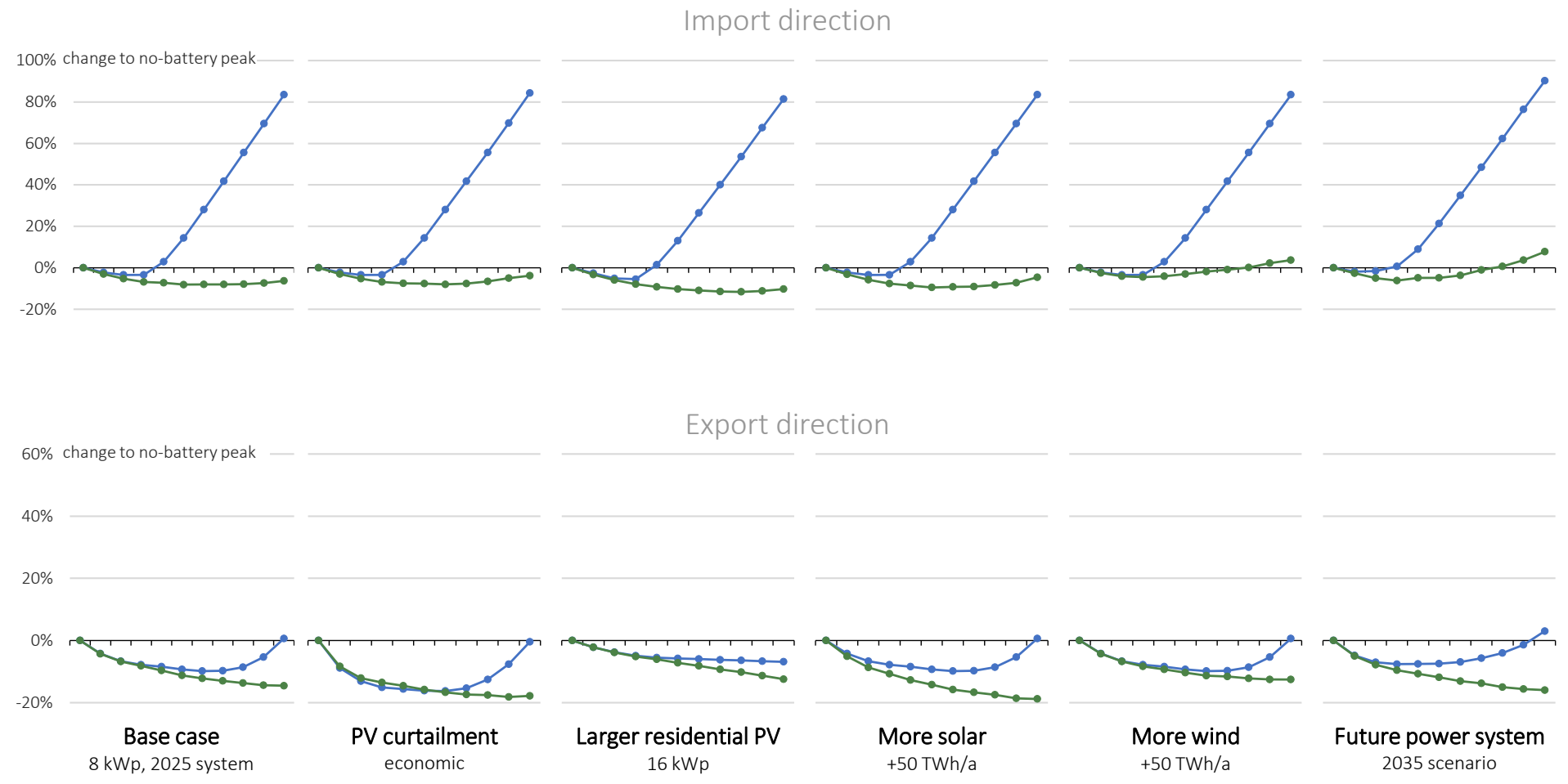

Figure 7. Coincident peak for exogenous and endogenous prices: renewables.

# 5.6. Network charges

In the base case, households pay a volumetric network charge of 0.10 EUR/kWh (including taxes and surcharges). Injection is free of charge. Batteries that reinject stored energy previously drawn from the grid receive a refund of the charge. I test these assumptions by halving and doubling the network charge and dropping the charge refund (Table A5).

Higher or lower charges do not change the results substantially (Figure 8). Requiring batteries to pay the full network charge reduces charging from the grid and thus the new import peaks. But in this case, too, endogenizing prices reduces the import peak substantially.

Two further variations change the structure of the network charge. Under net metering, which is common in the US, the network charge of 0.10 EUR/kWh applies only to each household's monthly net withdrawal: electricity fed in and drawn back within the month is not charged. Generation still reduces the charge, but shifting it in time no longer does. This removes the self-consumption motive for the battery and makes grid arbitrage cheaper. Under endogenous prices, the import peak then rises slightly above its no-battery level; under exogenous prices, both peaks rise about as much as in the base case. A network charge of 0.10 EUR/kWh on injection changes little.

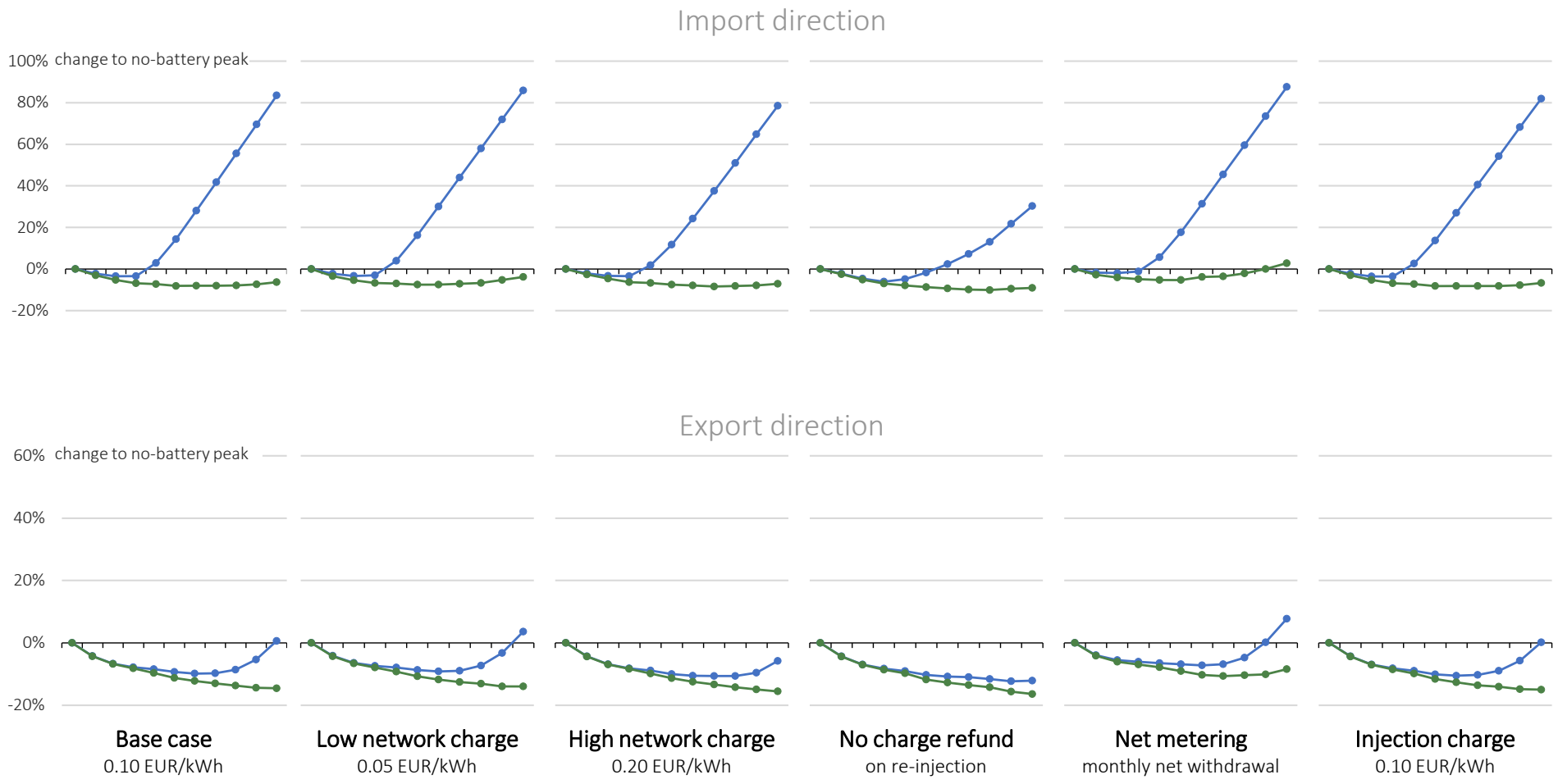

Figure 8. Coincident peak for exogenous and endogenous prices: network charges.

# 5.7. Electric vehicles

From a power system perspective, electric vehicles are batteries on wheels and can be optimized just like stationary storage. However, they are also net consumers of electricity. Their optimization is more constrained. As a first exploration, price-optimized EVs were added instead of stationary batteries in one sensitivity run; a detailed analysis is provided in Hirth (2026b). Specifically, EVs were randomly added to up to 50 % of households, with and without PV. Each EV has a capacity of 60 kWh and a charging power of 11 kW and is plugged in whenever parked at home. Individual driving profiles have a median distance of 14,000 km per year and 80 % of charging at home, which results in an annual electricity consumption of 1.8 MWh per car, increasing the local electricity consumption by 17 %. Discharging car batteries (vehicle-to-grid) is allowed; cycling costs and round-trip efficiency are the same as for stationary batteries.

At the same adoption rate, EVs increase both the import and export coincident peaks much more strongly than home batteries. This is driven by the higher power and capacity of EVs. For EVs, endogenous prices also reduce peaks substantially compared to exogenous prices. At 50 % of households with an optimized EV, exogenous prices increase the peak by 324 % over the no-battery peak; with endogenous prices, the increase is 24 %.

# 6. Robustness

This section checks the robustness of the price-feedback bias against three types of assumptions: the functional form of the price function, the measure of the coincident peak, and technical simulation procedures such as the draw of households. The findings persist across most of these variations. Section 6.4 and Figure A4 in the Appendix summarize the coincident peak at full adoption across all sensitivities, for the import and the export direction.

## 6.1. Alternative price functions

This paper is about wholesale prices responding to battery dispatch, so the most important robustness checks concern the price function. The base case models prices as a cubic function of net load estimated on historical prices, with the regression residual preserved, as in Equation (4). Here, this is compared to a range of alternative price functions:

- A quadratic function, see Equation (5)
- A piecewise linear function
- The cubic function without random price variation ($\varepsilon = 0$)
- A flatter and a steeper variant of the cubic function ($\alpha$ halved/doubled)

In addition to observed historical 2025 prices, I use each of the six price functions to create a further exogenous price curve, calculated for the case of zero batteries and then held fixed. This allows a direct comparison between exogenous and endogenous prices for each price function.

The coincident import peak varies little across price functions. Across all variants, the endogenous peaks stay below the no-battery level and far below the peaks under exogenous prices. For the export peak, the alternative functions lead to an even lower peak than the base case (Figure 9 and Table A5).

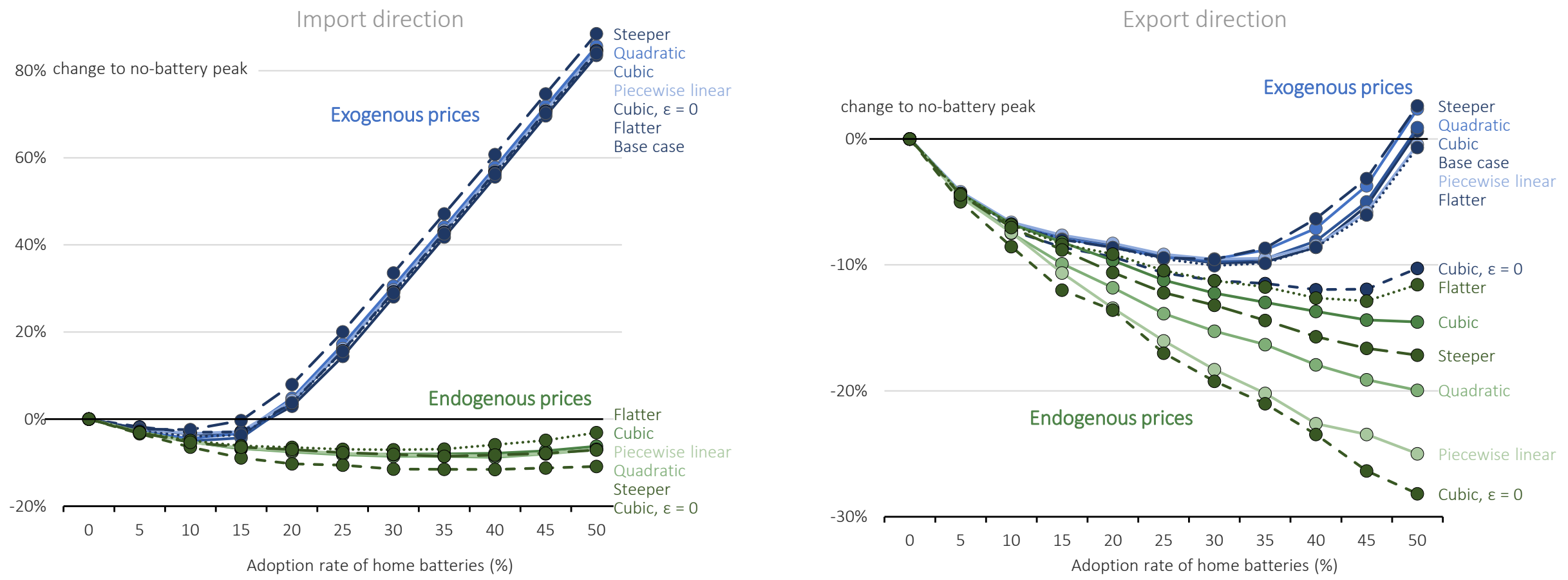

Figure 9. Coincident peak for alternative price functions and different exogenous price curves.

## 6.2. Alternative measures of the coincident peak

The next robustness check concerns the measure of the coincident peak. In the base case, I use the average load in the top 1 % of hours. Figure 10 and Table A5 show two additional metrics: the peak hour (highest loading of any hour) and the top 10 % of hours. The results hold for all three measures: compared to exogenous prices, the import peak at full adoption is reduced by 35–49 %, the export peak by 15–24 %.

Figure 10 also shows the coincidence factor: the coincident peak divided by the sum of the individual households' peaks, both measured as the mean of the top 1 % of hours. Under exogenous prices, it rises by about two thirds in the import direction; under endogenous prices, it falls slightly. In the export direction, it hardly changes under exogenous prices and falls under endogenous prices.

The 4-hour peak, the highest mean load over four consecutive hours, rises much less than the shorter measures under exogenous prices, because synchronized battery charging lasts only one or two hours; under endogenous prices, it stays at its no-battery level. Overload hours, the hours above the no-battery peak hour, rise to about 300 per year under exogenous prices and stay close to zero under endogenous prices.

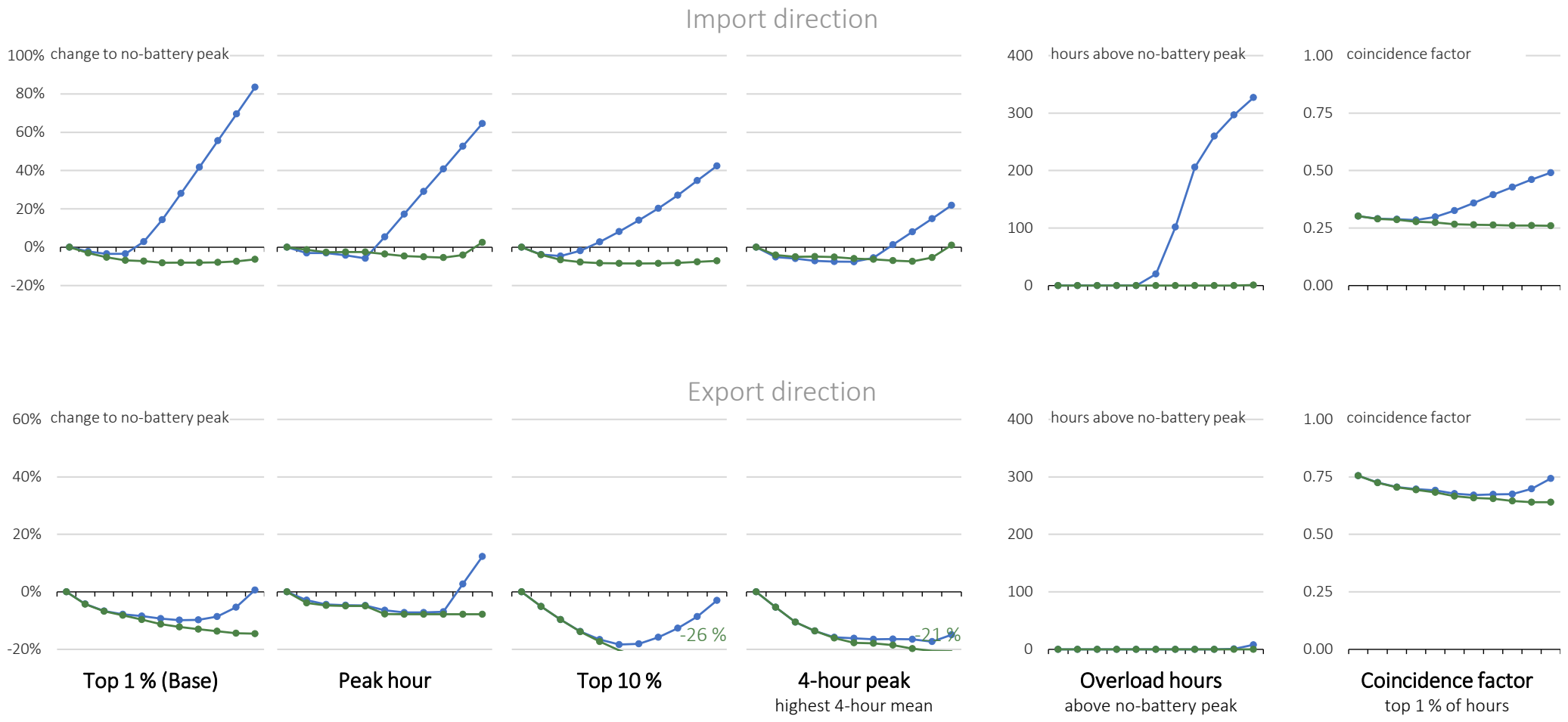

Figure 10. Alternative measures of the coincident peak (maximum, top 1 %, top 10 %, 4-hour peak), overload hours and coincidence factor for exogenous and endogenous prices.

# 6.3. Household simulation assumptions

The details of the simulation methodology may also affect results: the draw of households, the size of the household population, the coupling of household load to national load, measured (as opposed to simulated) household load profiles, the foresight horizon, the temporal resolution, the historical year, and PV orientation and share. Table A5 reports a number of robustness tests in this respect.

In every case, the coincident peaks at full adoption are lower under endogenous than under exogenous prices (import by 28–51 %, export by 9–25 %); they are also lower than without batteries, except with measured load profiles in the import direction.

# 6.4. Overall robustness

Figure 11 reports the coincident import peak at full battery adoption under exogenous and endogenous prices for all model runs. Across these runs, endogenous prices substantially reduce coincident peaks. The differences in coincident peaks are dozens or hundreds of percentage points in most cases. Also, while simulations with exogenous prices find peaks to increase by multiples in certain cases, with endogenous prices the increase is never larger than 25 %, something that is easily absorbed by the thermal margins of most distribution transformers. In most cases, price-optimized batteries reduce the coincident peak. Rather than fearing new peaks from price-optimized batteries, one may hope for the opposite.

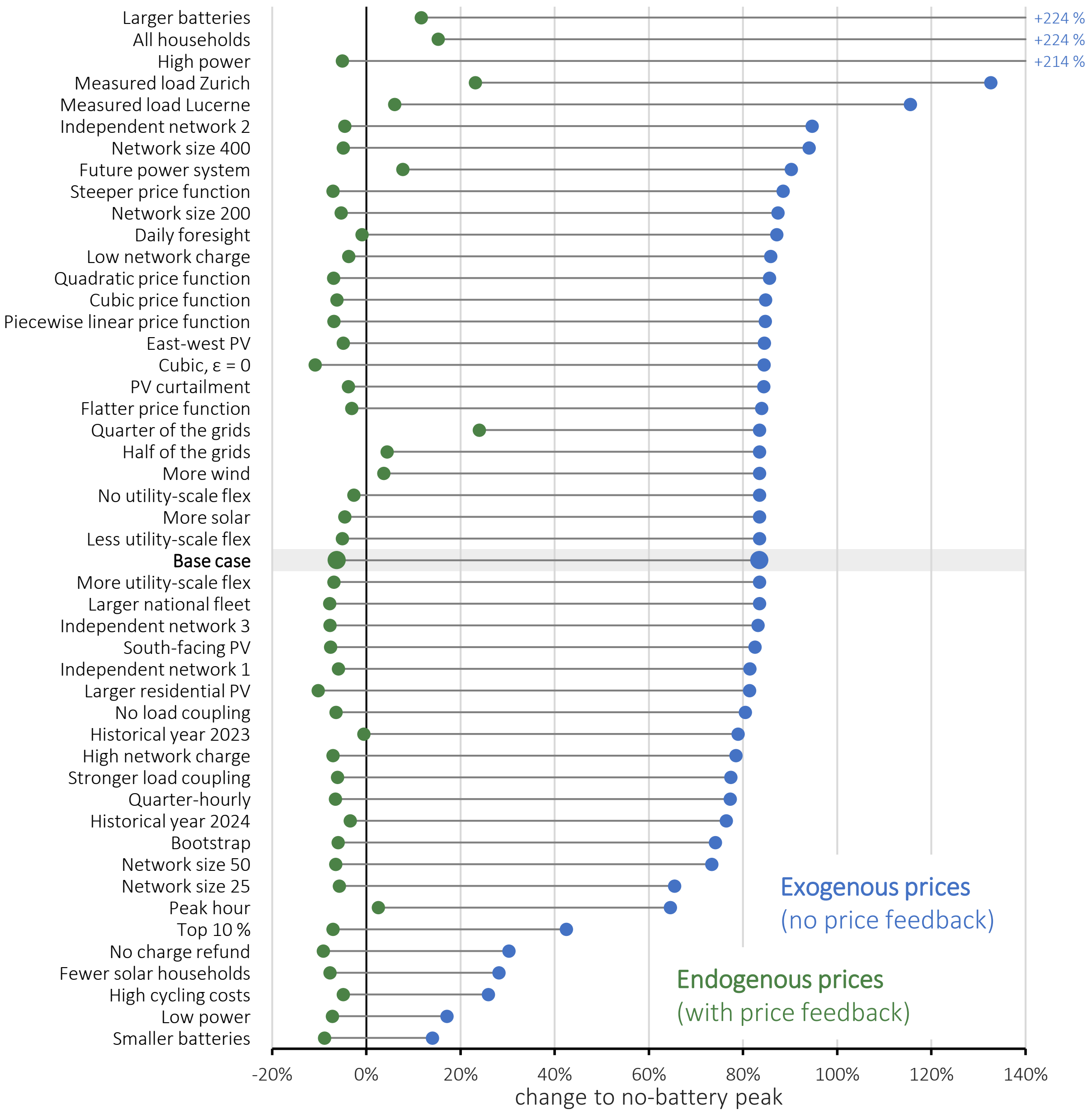

Figure 11. All model runs: endogenous (green) vs. exogenous (blue) prices at full battery adoption.

# 7. Analytical discussion

This section uses a simple analytical model to understand the simulation results in more depth. The model identifies the main parameters that determine whether price-responsive home batteries create a new local import peak under endogenous prices and how large it becomes. Hirth (2026a) provides a generalization.

## 7.1. Setup and assumptions

Consider a local distribution grid behind one transformer with $H$ home batteries, each with capacity $\overline{E}$ and rated power $\overline{P}$ (see Table 4 for symbols). The load of the grid without batteries, net of local PV generation, is the background load $B_t$. Its annual maximum $B_{max}$ is the old import peak, typically on a winter evening. The coincident load at the transformer is the background load plus the charging $\Lambda_t$ of the local battery fleet:

$$C_t = B_t + \Lambda_t \tag{6}$$

I am interested in determining if a new peak occurs. The model describes a single "critical night", the one where a new peak is most likely to occur when batteries charge during a period of low wholesale prices. This is the case if the maximum of $C_t$ exceeds $B_{max}$. To derive a closed-form solution, I have to apply more restrictive assumptions than in the simulations:

- National net load before flexibility, $R_t^0$, falls linearly to its minimum $R_{min}$ at hour $t_0$ and rises linearly afterward, with the same gradient $r$ on both sides ("V-shaped net load").
- The wholesale price has no random component ($\varepsilon = 0$).
- The retail price spread is large enough for every battery to charge from the grid and be fully charged at the end of the night.
- All flexible assets in the market (home batteries in all grids, utility-scale storage, import/export and other flexibility) follow the same arbitrage logic.

Table 4. Symbols of the analytical model

| Symbol | Definition | Unit |
|---|---|---|
| $t, t_0$ | Hour; hour of the lowest net load | h |
| $B_t, B_{max}$ | Background load of the local grid (load without batteries, net of PV); its annual maximum, the old import peak | kW |
| $B_{t_0}$ | Background load of the local grid in the hour of the lowest net load | kW |
| $\Lambda_t, \Lambda_{t_0}$ | Charging of the local battery fleet, $\sum_i c_{i,t}^h$; its value in the hour of the lowest net load | kW |
| $C_t, C^*$ | Coincident load at the transformer; its maximum, the new peak | kW |
| $H, \overline{P}, \overline{E}, \eta$ | Number of batteries in the grid; rated power, capacity and one-way efficiency of a battery (as in Appendix A4) | –; kW, kWh, – |
| $R_t^0, R_{min}$ | National net load before flexibility, $L_t - W_t - S_t$ (Eq. (2) without battery dispatch); its minimum in the night | GW |
| $r$ | Gradient of national net load | GW/h |
| $D$ | Peak national charging: charging of all flexible assets in the hour of the lowest net load | GW |
| $E$ | National flexibility: energy charged by all flexible assets during the night | GWh |
| $\sigma$ | Share of the local grid in national flexibility, $H\overline{E}/E$ | – |

## 7.2. The condition for new peaks

Under endogenous prices, a battery that charges in the cheapest hour raises the price in that hour. In equilibrium the price is flat across the hours in which the fleet charges. Because the price is strictly increasing in net load, a flat price implies flat net load after flexibility: charging raises net load to a constant plateau in these hours. Let $D$ denote the distance between the lowest net load and the plateau. It is the peak power at which the collective national flexibility charges, i.e., peak national charging. With V-shaped net load, net load before flexibility is

$$R_t^0 = R_{min} + r \cdot |t - t_0| \tag{7}$$

The entire energy that is charged forms a triangle with height $D$ and base $2D/r$. The area of the triangle equals the energy charged, which gives:

$$\frac{D^2}{r} = E \quad \Rightarrow \quad D = \sqrt{rE} \tag{8}$$

The price function does not appear – it affects the price level, but not the physical quantities. Each grid charges its share $\sigma$ of national charging ($-b_t$ in Eq. (3)) in every hour, so that the local charging peak occurs at the lowest net load and equals the share times peak national charging:

$$\Lambda_{t_0} = \sigma \cdot D = \sigma\sqrt{rE} \tag{9}$$

The new coincident peak is the background load in the hour of the lowest net load, $B_{t_0}$, plus the local charging peak:

$$C^* = B_{t_0} + \sigma\sqrt{rE} \tag{10}$$

A new peak emerges if the local charging peak exceeds the headroom at the lowest net load:

$$C^* > B_{max} \quad \Leftrightarrow \quad \sigma\sqrt{rE} > B_{max} - B_{t_0} \tag{11}$$

Four parameters determine the new peak:

- The background load of the local grid in the hour of the lowest national net load (usually 2–3 a.m.). The higher the background load is at that moment in time, the more likely it is that batteries that charge push the load above the previous peak.
- The share of the local grid in national flexibility (larger if there is a local cluster of home batteries, smaller if there is abundant pumped hydro or utility-scale battery capacity). The more concentrated flexibility is, the more likely a new peak emerges.
- The amount of national flexibility itself (how much electricity can be absorbed). For a given distribution of flexibility, a larger total capacity means that more energy is absorbed, and that peak charging capacity increases.
- The gradient of net load. If cheap electricity is concentrated within a few hours, batteries charge at a higher power.

It is worth noting that rated power does not appear as a parameter that determines the new peak. Rated power does not bind as long as the effective charging duration $\sqrt{E/r}$ exceeds the full-load duration $\overline{E}/(\eta\overline{P})$ of a battery. The level, slope and shape of the price function, the network charge and the arbitrage wedge do not enter. They determine whether batteries charge from the grid, not how high the peak is. Market size does not enter either: if a market is split into identical zones, the share of a grid rises by the same factor by which peak national charging falls.

## 7.3. Comparison with the simulations

Figure 12 illustrates the national electricity market during the critical night (top) and the resulting distribution grid transformer load (bottom) using empirical parameters of the night from 9 to 10 December from the same dataset used for the simulations (Table 5), with $E = NH\overline{E}$ and $\sigma = 1/N$.

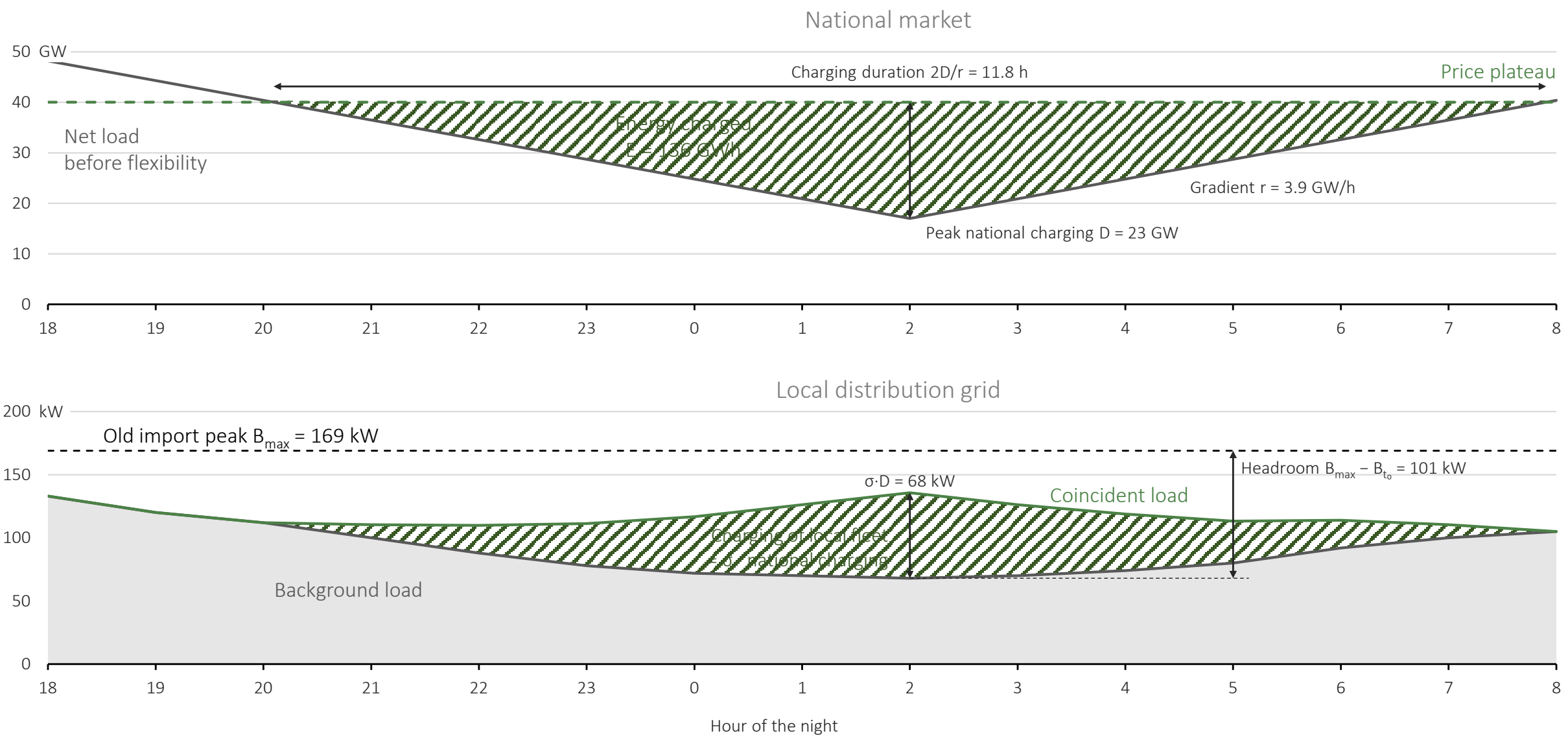

Figure 12. National net load on the critical night with the energy charged by the national fleet under V-shaped net load (top) and local import load (bottom).

These parameters can be inserted into Equations (9) and (10) to determine the local charging peak and the new coincident peak. The numbers are within 3 % of the simulation results. The good fit suggests that the stricter assumptions of the analytical model are not introducing significant biases.

Table 5. Analytical model and simulation on the critical night of the base case

| Symbol | Quantity | Analytical model | Simulation |
|---|---|---|---|
| $H, \overline{P}, \overline{E}$ | Number of batteries in the grid; rated power and capacity of a battery | 50; 4 kW, 8 kWh | |
| $B_{max}$ | Old import peak | 169 kW | |
| $B_{t_0}$ | Background load in the hour of the lowest net load | 68 kW | |
| $r$ | Gradient of national net load | 3.9 GW/h | |
| $E$ | National flexibility (home battery fleet), $NH\overline{E}$ | 76 GWh | |
| $\sigma$ | Share of the local grid in national flexibility, $H\overline{E}/E = 1/N$ | $5.3 \times 10^{-6}$ | |
| $D$ | Peak national charging | 17 GW | – |
| $\Lambda_{t_0}$ | Local charging peak | 91 kW | 93 kW |
| $C^*$ | New coincident peak | 159 kW | 163 kW |

# 8. Conclusions

This paper asks whether the new grid peaks attributed to price-optimized residential batteries in previous studies persist when wholesale prices respond to battery dispatch. It finds that in most cases they do not.

With exogenous prices, the simulations reproduce the established finding of the literature: with batteries often (dis)charging in perfect synchrony, once more than 15–20 % of households install a battery at home, they create new peaks in what used to be off-peak periods. These peaks can be large; under my central assumptions, the import peak rises by 83 %. With endogenous prices, however, the same fleet *reduces* the import peak by 6 %. The export peak, which occurs when local solar generation exceeds demand, is also reduced.

However, the price feedback does not rule out new local peaks. Wholesale prices clear at the level of the bidding zone (or the transmission node), not of the local distribution grid, so a local fleet can still be synchronized by the national price whenever the local situation strongly deviates from the national average. This is the case, e.g., when certain affluent suburban neighborhoods adopt batteries much faster than the country as a whole. Nevertheless, these results suggest that published studies overstate the local peak-load effects of price-responsive batteries because they ignore the feedback of batteries on wholesale prices.

This "price-feedback bias" matters for grid planning. Reinforcement decisions that rest on exogenous-price simulations overstate the transformer and cable capacity that price-responsive batteries require. Future work could quantify the resulting investment in unnecessary capacity.

The feedback also matters for regulatory and policy decisions. It suggests that dynamic pricing is not only beneficial for reducing generation costs but also for the local grid. This reduces the need for new dedicated instruments to address grid constraints, such as limits on charging power or time-varying network charges.

This study simulates stationary home batteries. I expect the same effects to occur with other types of residential flexibility, such as price-optimized electric vehicle charging (see Hirth, 2026b), heating, and cooling. The simple price feedback method proposed here could be applied in future studies covering a wider range of technologies.

# Appendix

## A1 Findings of previous studies

Table A1. Studies on electric vehicles and other flexible loads

| Study | Setting | Signal | Grid metric | Synchronized new peak? | Price feedback |
|---|---|---|---|---|---|
| Turk et al. (2025) | Massachusetts; 400 households on one feeder; EVs with perfect foresight | TOU energy rate, two-part, plus seven network tariff designs | Feeder peak (import) | Yes. Rebound peak at the start of the off-peak period exceeds the historical evening peak from 15 % EV adoption; at 50 % adoption it is 60 % above the historical peak | No |
| Kamana-Williams et al. (2025) | Auckland, New Zealand; 50 households on one transformer | TOU retail price with one or two staggered off-peak windows | Transformer peak (import) | Yes. Secondary peak at the start of the off-peak period; it exceeds the original peak from 20 % participation and by 33 % at full participation | No |
| Gottwalt et al. (2011) | Germany; 20 populations of 1,000 single-family households with smart appliances (refrigerator, freezer, washing machine, dishwasher, dryer, electric water and space heating); no grid model | RTP: day-ahead hourly rate from the 2008 EEX spot price; smart-appliance adoption 0 to 90 %, price spread varied by ±10 % | Aggregate peak of 1,000 households, 100 highest hours (import) | Yes. In every scenario with load shifting the 100 highest hours exceed the flat-rate peaks; the highest hour rises from about 1.1 to about 1.4 MW. New peaks form where off-peak pricing sets in and grow with the adoption rate of smart appliances (avalanche effect) | No |
| Verzijlbergh et al. (2014) | Netherlands; one distribution cable with about 1,000 households and 500 EVs | RTP: wholesale prices from a Dutch system model | Feeder peak (import) | Yes. Cost-minimizing charging raises the cable peak by 47 %, more than uncontrolled charging (35 %); with 2 GW wind the new peak does not occur. Static time-of-use network tariffs raise it by up to 69 % | Yes |
| Stawska et al. (2021) | Netherlands; 250 households on one 500 kW feeder; EV aggregator | Other: historical Dutch imbalance prices; energy tariff, capacity tariff or local flexibility market | Feeder overload hours | Yes. Charging spikes whenever imbalance prices are very low, growing with adoption and charging power; an energy-based network tariff does not prevent it | No |
| Flath et al. (2014) | Germany; EV fleet; residential distribution grids | Other: time-based prices, uniform price vs. area pricing with a local capacity component | Distribution grid peak (import) | Yes. Time-based prices alone produce high load spikes from simultaneous charging, independent of charging strategy and power level | No |
| Holmberg et al. (2023) | US; IEEE 8500 reference feeder with 1,977 single-family homes, each with an air-source heat pump; 8 MW peak load | RTP: CAISO five-minute real-time prices, against hourly averages as a day-ahead proxy | Feeder loading (import); voltage; voltage violations | Yes, as synchronized swings rather than as an annual peak. Under five-minute prices the mean change in substation flow between time steps rises from 101 to 870 kW, about 10 % of peak load. Voltage violations rise to about 400 % of the baseline | No |
| Huang and Sandström (2025) | Sweden; residential district with 1,321 customers on twelve 10/0.4 kV substations; EVs with V1G and V2G at 5 to 70 % adoption | RTP: spot price, plus a time-of-use network tariff; also prices differentiated by substation | Transformer loading; cable loading; voltage; hosting capacity | Yes, above 20 % adoption. V2G lowers peaks at low adoption and raises them above it. Prices differentiated by substation lower the district peak but not the peak of the individual substation | No |
| Salah et al. (2015) | Switzerland; 49 distribution substations of one regional grid with forecast 2040 peak-day load; survey-based driving profiles; home charging at 11 kW; adoption of 16, 50 and 100 % | RTP: hypothetical dynamic rate following the Swiss spot price, against a flat rate | Substation loading; share of substations overloaded (import) | Yes, with medium and high renewable shares: overloads from 16 % adoption despite the price feedback. With low renewable shares fewer overloads than under the flat rate, which overloads 6 of 49 substations at full adoption | Yes; spot price as a function of net load, updated after each vehicle's charging decision |
| von Bonin et al. (2022) | Germany; 100 metropolitan households, each with EV, PV and 11 kW wallbox; home energy management; 2030 prices | RTP: hourly day-ahead prices for 2030 from a market simulation | Aggregate peak, 100 households (import) | Yes. Price-optimized charging raises the aggregate peak by 184 % over household load, direct charging by 84 %; the relative increase grows with the number of households | No |
| Panda et al. (2024) | Netherlands; measured sessions at about 1,200 public charging points, 2022; aggregation of 1 to 1,024 points | RTP: Dutch day-ahead prices of 2022; variants with segmented network tariffs | Aggregate peak per charging point; diversity factor (import) | Yes. Day-ahead-optimized charging raises the average peak by up to 16 % over unoptimized charging | No |
| Flath and Gottwalt (2016) | Germany; microgrid with 900 households, EVs and wind generation; no network model | Other: posted price as a linear function of local net wind; TOU derived from it | Aggregate peak reserve import, 900 households | Yes. Under open-loop prices the reserve peak is about 11 times that under a flat rate; a closed-loop price, updated after each charging decision, removes it | Yes, in the closed-loop variant |
| Hennig et al. (2022) | Low-voltage neighborhood with 50 German household profiles, up to 25 EVs, 120 kW transformer | RTP: day-ahead prices of 2020, with fixed, volumetric, capacity subscription or mixed network tariffs | Transformer loading (import) | Yes. Under fixed and volumetric network tariffs the transformer is overloaded from about 8 EVs; a capacity subscription avoids it | No |
| Jones et al. (2022) | US; ten distribution feeders; EV adoption for 2030 | TOU: on-peak period noon to 7 p.m.; immediate or randomized charging start | Feeder peak (import) | Yes. Immediate charging start raises the residential feeder peak by 20 %, with a new peak at 8 p.m.; a randomized start lowers it by 5 % | No |
| FTI Consulting (2025) | EU; 200 households on one feeder, EV adoption 0 to 60 % | RTP: dynamic energy rate, with four network tariff designs | Aggregate feeder peak (import) | Yes, above 15 % EV adoption under a flat volumetric network tariff; capacity-based and dynamic network tariffs cut the peak at 60 % adoption by 8 to 24 % | No |
| Heinrich et al. (2021) | Denmark (Bornholm); MV feeder, 169 EVs and 150 heaters; transformer rating reduced to 1,200 kVA | RTP: day-ahead prices, aggregator optimization; capacity limitation service | Primary substation transformer loading (import) | Yes. Night charging at low prices creates off-peak congestion with up to 825 kW of flexible consumption | No |
| Ziras et al. (2025) | Denmark (DK2); 200 residential 11 kW EV chargers, charging data 2022 to 2023 | RTP: spot price plus time-of-use network tariff, aggregator cost minimization | Aggregate load of 200 chargers (import) | Yes. Load concentrates between midnight and 3 a.m., up to about 1,100 kW, against a historical peak of about 250 kW for 200 households | No |
| Unterluggauer et al. (2023) | Denmark (Frederiksberg); three meshed urban low-voltage grids with 213 to 562 households; 40 % of cars electric, 11 kW home charging; Monte Carlo power flow for a winter weekday | RTP: hourly spot price plus time-of-use network tariff, cost-minimizing EV charging; time-of-use network tariff alone | Transformer and cable overload probability; peak load (import) | Partly. Charging synchronizes at night, but the peak exceeds that of uncontrolled charging in at most 1.4 % (time-of-use) and 0.5 % (spot price) of runs. If charging is optimized over several days and forced into the same night, new peaks occur in all runs, with transformer loading above 150 % | No |
| Li et al. (2014) | Denmark; RBTS bus 4 test distribution system with 4,779 customers; EV aggregators with Danish driving data; EV penetration 100 and 500 % of the household peak | RTP: system locational marginal price; distribution locational marginal prices as alternative | Line loading of the medium-voltage feeder (import) | Yes. Under the system price alone, aggregators charge all EVs in the lowest-price hour and congest the feeder; distribution locational prices spread charging over several hours | No |
| Hennig et al. (2024) | Netherlands; neighborhood of 50 households with 25 EVs; two weeks of January 2021, hourly | RTP: EPEX day-ahead prices, retailer-optimized charging; capacity limitation contracts and capacity subscription as alternatives | Aggregate load of the neighborhood (import) | Yes, qualitatively. Price-optimized charging produces the highest load spikes in low-price hours; capacity limitations spread the load | No |
| Wu et al. (2025b) | South Australia; urban feeder with 56 distribution transformers and 2,059 residential customers; EV adoption 50 and 100 %; one year of smart meter data | TOU: off-peak from 1 a.m., adopted by 50 or 100 % of EV owners; staggered off-peak start times as alternative | Transformer utilization; overloaded transformers (import) | Yes. Charging at the start of the off-peak window creates an early-morning peak more than 3.5 times the EV peak under a flat rate; at 100 % EVs and full TOU adoption 34 of 56 transformers overload, against 5 under the flat rate; staggered start times avoid overloads | No |
| Hein et al. (2026) | Germany; SimBench rural low-voltage feeder with 97 buses and a 250 kVA transformer; EVs (60 kWh, 11 kW), heat pumps and PV at 0 to 90 % adoption; 15-minute resolution, one week per season | RTP: exogenous 15-minute intraday prices; variants with a time-variable network charge (§14a module 3) and transformer-aware central schedules | Transformer and line loading; voltage (import) | Yes. Price-optimized scheduling raises transformer loading to 253 % of rating at 30 % adoption, against 38 % without EVs and heat pumps and 112 % with random charging start; 478 % at 90 % EV adoption. The time-variable network charge does not lower winter peaks | No |
| Lundblad et al. (2026) | Sweden; synthetic low-voltage grids, about 105,000 grid cells; home charging of 25 to 125 % of the vehicle fleet | RTP: hourly spot prices, alone or with a monthly peak power network tariff; logged charging as reference | Peak power; transformer and cable overloads; voltage violations (import) | Yes. At full fleet electrification, average low-voltage peak power rises by 73 % with spot-price charging, 51 % with logged charging and 10 % with the peak power tariff, against no EVs; with spot-price charging, 60 % of grid cells have violations from 75 % electrification | No |
| Jung and Sundström (2023) | Sweden; real low-voltage grid with 30 customers (Linköping); every household with 10 kWp PV and an EV, share of EVs at home in daytime 0 to 100 %, charging power 3 to 22 kW, mostly vehicle-to-grid; Monte Carlo simulation | RTP: spot prices of October 2020, with and without a monthly peak-power network charge on the import peak or the absolute peak | Voltage variation (import and export) | Yes. Voltage variation rises markedly with charging power under spot-based rates; synchronized selling by many households moves the voltage peak to other hours instead of flattening it | No |
| Wu et al. (2025a) | South Australia; 11 kV feeder with 56 distribution transformers, 2,059 residential and 60 commercial customers; year-round smart meter data; EV adoption up to 80 %; TOU adoption 0 to 100 %; mixed charging power and frequency | TOU: existing retail time-of-use rate, charging timed to the start of the off-peak period (around 1 a.m.), against a flat rate with charging on arrival | Maximum transformer utilization; unserved energy (import) | Yes. Timers synchronize charging at the start of the off-peak period: the peak of a 1,000-EV fleet is 3,688 kW against 871 kW under the flat rate. At 80 % EVs and 25 % TOU adoption, peak-day transformer utilization falls from 133 to 125 %, but on average and high-export days the off-peak spike raises maximum utilization | No |
| Shrestha et al. (2025) | Maryland, US; seven representative feeders clustered from 1,500 feeders of one utility; smart charging enrollment 0 to 50 % by 2035 | TOU: charging paused from 5 to 9 p.m.; load balancing as alternative | Feeder peak; transformer overloads; upgrade cost (import) | Yes. At 50 % enrollment, TOU cuts feeder peaks by 0 to 12 %; at high enrollment the peak moves to the end of the TOU window and forms a second peak, on one feeder raising transformer upgrade cost | No |
| Elmallah et al. (2022) | Northern California, US; 3,043 circuits and 724 substations of one utility; 3.1 million EVs in 2030 rising to 12.5 million in 2050, plus electrified heating | TOU: rate dropping at midnight (base case); spread nighttime charging and daytime charging as alternatives | Feeder and substation capacity exceedance; upgrade cost (import) | Yes. The midnight price drop creates a midnight charging spike; 95 to 260 feeder upgrade projects per year until 2030, against 53 to 87 planned; cumulative upgrade cost 1 to more than 10 billion USD by 2050 | No |
| Unterluggauer et al. (2022) | Denmark (Frederiksberg); urban 10 kV network with 284 transformers and 440 cables; full electrification of the 2020 car fleet (24,252 resident and 34,818 visiting EVs), 11 kW home and workplace charging; 40 peak-load weekdays | Other: synchronized start signal: home charging delayed to midnight (stylized low-price period) or 6 p.m., workplace charging to 10 a.m.; against uncontrolled charging | Aggregate peak load; transformers and cables above 100 % loading; reinforcement cost (import) | Partly. With non-daily charging, synchronizing at midnight raises the aggregate peak by less than 32 %, like uncontrolled charging. If all vehicles charge daily or on the same day, the peak rises by 183 to 215 % and 108 to 178 of 284 transformers overload | No |
| Brockmann et al. (2025) | Norway (bidding zone NO5); agent-based model of 200 EVs calibrated on about 28,000 home charging sessions; year 2023 | RTP: day-ahead price plus time-of-use network charge, cost-minimizing charging in each session, with price-sensitive day-to-day plug-in decisions; TOU and flat variants | Coincidence factor of EV fleet charging for 10, 100 and 1,000 EVs, no household load (import) | Yes. Price-sensitive users plug in on the cheapest nights: for 100 EVs the coincidence factor rises from 31 % (no price sensitivity) to 46 % (high) and 54 % (maximum), against 20 % under a flat price; the diversity benefit of larger fleets almost disappears | No |
| Brinkel and van Montfoort (2026) | Netherlands; urban grid with 650 households and public charging at 380 stations (196,859 sessions of 2022), scaled to 0 to 100 % EV adoption; charge point operators optimize charging | RTP: Dutch day-ahead prices of 2022, with six network tariff designs; fixed retail price as alternative | Distribution grid expansion cost, driven by the aggregate 15-minute peak (import) | Yes. Under day-ahead prices charging concentrates in the cheapest hours with peaks above uncontrolled charging; at 100 % adoption grid costs exceed the no-EV level by more than 80 %, against 60 % under uncontrolled charging | No |
| Winzer and Hensler-Ludwig (2024) | Switzerland (Aargau); 300 households behind one transformer, synthetic profiles; EVs (11 kW), heat pumps, PV and PV-charged home batteries; device scenarios 2020 and 2050 (78 % of households with flexible devices) | TOU: network tariff with daily high- and low-tariff periods; CPP; Other: tariff proportional to forecast grid load; RTP: spot-proportional tariff | Aggregate peak, 300 households (import) | Yes. Time-of-use and critical peak tariffs raise the peak above the flat-rate peak by 5 to 15 % in 2020 and by more than 40 % in 2050; the grid-load tariff more than doubles it in 2050 | No |
| Iftikhar et al. (2026) | Pakistan (Lahore); 11 kV residential feeder with 60 transformers and 301 nodes; synthesized seasonal load; 23 or 46 EVs (5 and 10 % of feeder peak load) with 1.9 kW chargers and V2G | RTP: day-ahead hourly price from a national optimal power flow, with aggregator scheduling subject to transformer limits, against a flat rate and a two-part TOU rate | Transformer loading at weak nodes (import) | Partly. Hourly pricing keeps winter evening loading below the 80 % threshold where flat and two-part rates exceed it, but loading at one node rises in late-night low-price hours | No |
| Ademulegun et al. (2026) | United Kingdom (Northern Ireland); headroom analysis of 5,620 distribution transformers (100 to 1,000 kVA); one and two EVs per household | TOU: rate with night (1 to 8 a.m.) and midday (12 to 4 p.m.) off-peak windows, against charging in normal household demand hours | Share of distribution transformers needing upgrade (import) | No. Off-peak charging cuts the share of transformers needing upgrade from 67 to 52 % (one EV per household) and from 78 to 63 % (two EVs). Charging within the off-peak windows is not time-resolved, so synchronization at their start is not examined | No |
| Lilienkamp et al. (2025) | Germany; 16,191 representative low-voltage grids (402 NUTS-3 regions, six SimBench grid types, eight transformer sizes) aggregated as virtual storage into a European hourly dispatch model; 15 million EVs in 2030 and 35 million in 2045; home charging only, no V2G | RTP: hourly wholesale prices from the dispatch model (endogenous); benchmark run with the same price series fixed exogenously | Transformer loading (charging shifted to respect transformer capacity); transformer expansion need (import and export) | Yes. Price-optimized charging concentrates in low-price hours and adds up to 40 GW in single hours nationally; to respect transformer limits, up to 7.8 GW (2030) and 28.8 GW (2045) of charging must be shifted locally. With exogenous prices the national EV load peak is 75 % higher than with endogenous prices. Vehicle-driven transformer expansion need is 1.8 of 16 GW in 2030 (rooftop PV dominates) and 27.7 of 51.6 GW in 2045. | Yes; prices are the duals of a European dispatch model (linear program) with aggregated EV flexibility; comparison run with exogenous prices |
| Qu et al. (2026) | Stylized setting with a representative UK winter day, hourly; 1,000 EVs in 25 categories; share of price-responsive EVs 0, 30 and 50 %; with and without V2G; no grid model | Other: day-ahead retail price set by a profit-maximizing aggregator against an exogenous wholesale price (0.02 to 0.20 EUR/kWh, daily mean equal to the wholesale mean); against uncontrolled charging | Aggregate EV load, 1,000 EVs, no household load (import and export) | Partly. Price-responsive EVs charge in the low-price hours 2 to 5 a.m. With 30 % price-responsive EVs the night peak stays below the uncontrolled evening peak of about 960 kW (−25 % without, −7 % with V2G). With 50 % it exceeds it (+25 % without, +55 % with V2G). Values read from figures | No |
| Van Montfoort et al. (2026) | Netherlands; randomized trial with 600 households with EV chargers | Other: network charge on consumption above 5 kW, static or TOU, 5 p.m. to 2 a.m. | Household peak (withdrawal); aggregate peak, 10 households | Yes, under the time-of-use version. Charging moves to after 2 a.m. without lowering own peaks; on clusters the new violations outweigh the relief during the priced hours. The static charge has no cluster effect | No |
| Torriti (2012) | Province of Trento, Italy; 1,446 residential smart-meter customers on 41 substations; flat rate in year one against TOU in year two, controlled for temperature | TOU, two periods (peak 8 a.m. to 7 p.m., off-peak otherwise) | Substation demand during morning and evening peak events (import) | Yes. The morning peak shifts. The evening peak persists and a higher peak appears after 9 p.m., plus a third peak at 4 to 5 p.m.; 31 of 41 substations (76 %) see higher demand during peak events; consumption rises by 14 % | No |
| Brylle et al. (2026) | Denmark; metered hourly load of one residential secondary substation with a high EV share (one week, April 2025) and of industrial sites in the Cerius-Radius grid, 2023–2025 | RTP: Nord Pool spot price, plus a static time-of-use network charge | Transformer loading; exceedance events (import) | Yes. 76 % of all capacity exceedance events fall in the low-load night window, which is also the low-price window, with load up to 50 % above the transformer rating | No |
| Ovaere and Vergouwen (2026) | Flanders, Belgium; about 42,000 households with smart meters, staggered opt-in to RTP, difference-in-differences | RTP, combined with an individual peak demand charge | Household peak (withdrawal); aggregate peak, 10 and 25 households (import) | Yes, small. Household monthly peak up by 76 W; aggregate peak up by 1 kW at a high RTP share on 10-household circuits, not significant on 25-household circuits | No |
| Bailey et al. (2025) | Alberta, Canada; randomized field experiment with 202 EV owners, July to December 2023; EVs grouped on virtual transformers of 10 vehicles with capacities of 12 to 24 kW | TOU: reward of 0.035 EUR/kWh for home charging from 10 p.m. to 6 a.m. and 10 a.m. to 2 p.m.; managed charging as second treatment | Transformer capacity violations, virtual transformer with 10 EVs (import) | Yes. Off-peak violations rise by 133 % and total violations more than double, exceeding the pre-treatment peak; managed charging halves them | No |
| Becker et al. (2025) | Flanders, Belgium; 42,089 residential smart meter profiles (862 heat pumps, 1,924 EVs), statistically sampled feeders of 10 to 250 connections; 2022 | TOU: day/night rate | Sampled feeder peak (import and export); timing of the peak | Yes. Feeders with high-power EV charging peak between 10 p.m. and 1 a.m. in winter, which the authors attribute to charging under the night rate | No |
| Samende et al. (2025) | UK; charging data of 96 EV households in a living lab, managed by an aggregator; simulated semi-urban (120 households) and urban (205 households) networks with 100 to 500 kVA transformers; EV adoption 10 to 100 % | TOU: lowest price from 11:30 p.m. to 5:30 a.m.; aggregator-optimized charging | Transformer peak (import); voltage | Yes. From about 20 % adoption a new peak forms around 1 a.m.; above 30 % it exceeds the evening peak, by a factor of up to 2.4 (semi-urban) and 4.0 (urban) | No |
| Brinkel et al. (2023) | Utrecht, Netherlands; field experiment with public charging stations in five low-voltage grids | RTP: day-ahead prices, optimized by the charge point operator, plus a three-tier network charge by transformer loading | Transformer congestion time (import) | No. Congestion time falls by 21 %, from 1.6 to 1.3 % of the time, against uncontrolled charging; the low-price tiers are capped to avoid a delayed peak | No |

# A2 Literature search

The review draws on a pool of more than 360 papers. The pool combines papers collected by the author with the results of three searches conducted between September and November 2026. First, keyword searches on the web covered home batteries, electric vehicles and other flexible household loads under time-varying prices and their effect on distribution grids. Second, gaps were closed by following the references of the studies found. Third, a forward citation search in OpenAlex in October 2026 started from 68 core studies and returned 3,139 citing works published since 2020. It was complemented by ten keyword searches on endogenous prices, rebound peaks, synchronization, price feedback, virtual power plants and vehicle-to-grid.

All papers were screened, in full text where available. A study is included in Table 1 or Table A1 if it meets three criteria. First, flexible household assets are dispatched in response to a time-varying price, such as a time-of-use or real-time retail rate. Second, the study reports a measure of stress on the local grid, such as coincident peak load, transformer or feeder loading, voltage, or reinforcement cost. Third, this measure refers to a population of households rather than to an individual household. Table 1 covers home batteries. Table A1 covers electric vehicles and other flexible loads.

The column “Price feedback” reads “Yes” only if households respond to prices and prices respond to the aggregate response of households.

Table A2 lists studies that were reviewed in detail but excluded, with the reason for exclusion.

Table A2. Studies reviewed but excluded

| Study | Subject | Reason for exclusion |
|---|---|---|
| Kühnbach et al. (2021), Powell et al. (2022) | Price-driven and timer-driven charging of national electric vehicle fleets | System level: the metric is the national net load or net demand peak. No distribution grid is modeled. Kühnbach et al. (2021) model the price feedback |
| Ensslen et al. (2018), Martin et al. (2025) | A load-shift rate in an agent-based market model, and emissions-based charging signals | No grid metric; the outcomes are hours of capacity deficit and wholesale prices, and grid emissions. Both model the feedback of the fleet on the signal, at system level |
| Signer et al. (2024) | Vehicle-to-grid trading under four tax and levy regimes in an agent-based German day-ahead market model to 2040 | No grid metric, system level; the outcomes are vehicle-to-grid revenues, wholesale prices and storage investment. The study models the feedback of the fleet on wholesale prices |
| Stute et al. (2024) | Dynamic retail rates for prosumers with electric vehicles and heat pumps | No grid metric; the study reports the timing of household load only |
| Verzijlbergh et al. (2012) | Controlled EV charging on 55 Dutch distribution networks | No price signal; the controlled case minimizes the combined network peak on behalf of the distribution utility |
| Figgener et al. (2024), Zhou et al. (2021), Zhou et al. (2022) | Observed home storage fleets in Germany and Australia | The systems run for self-consumption on flat rates, not against a time-varying price. Figgener et al. (2024) also report no grid metric |
| Askeland et al. (2025) | Flexibility of energy communities in Norwegian distribution grids, in an equilibrium model | The signal is a local market price and direct procurement of flexibility by the distribution utility, not a time-varying retail price derived from the wholesale market |
| Gupta and Morey (2023) | Demand response trials with heat pumps and batteries in 14 UK dwellings, measured at the substation feeder | No price signal; the interventions are dispatched directly by the trial operator. The feeder effects are large, up to 307 % in the turn-up direction |
| Lorenz et al. (2026) | Dynamic against fixed retail rates for PV-battery households, simulated on 448 metered German households | No grid metric; the study reports household gains only |
| Priyadarshan et al. (2025) | Distribution grid capacity needed for full residential electrification across US counties | No price signal; load is not price-responsive and reinforcement is costed at static rates |
| Stute and Klobasa (2024) | Time-of-use and real-time pricing with four network charge designs on six SimBench low-voltage grids | No no-flexibility baseline; the study compares network charge designs against each other; transformer loading of up to 339 % of rating is reported for the tariff variants only |
| Patteeuw et al. (2015), Arteconi et al. (2016) | Integrated modeling of active demand response with electric heating in a stylized Belgian system, with centrally controlled demand, the supply side and the demand side solved jointly, and the effect of participation rates | No grid metric and no retail price; the outcomes are system operating cost and savings per participant. The studies are the closest precedent to the argument made here: with an exogenous price profile the demand side overreacts and creates large peaks in low-price hours (Patteeuw et al.). The benefit per participant falls as participation rises (Arteconi et al.) |
| Klaassen et al. (2016) | Field test of dynamic rates with smart appliances and home displays in a Dutch smart grid pilot | No storage or vehicles; the flexible load is white goods only; the group peak of 77 and 111 households does not respond to the timing of peak prices |
| Schram et al. (2018), Moshövel et al. (2015) | Peak shaving with PV-battery systems sized for self-consumption in 79 Dutch households, and grid relief from forecast-based charging of PV-battery systems in Germany | No price signal; the batteries run for self-consumption or follow a peak-shaving rule with forecast or perfect foresight. Schram et al. report a neighborhood peak reduction of 6 % without control and 51 % with perfect foresight |
| Crozier et al. (2020), Mangipinto et al. (2022) | Smart charging of electric vehicles in Great Britain, and mass deployment of electric vehicles with smart charging in 28 European countries | No price signal; charging is centrally optimized (Crozier et al.) or follows fixed charging rules (Mangipinto et al.). Mangipinto et al. report national peak demand only. Crozier et al. find that smart charging cuts the share of distribution networks needing reinforcement from 28 to 9 % |
| Fischer et al. (2019) | Stochastic bottom-up model of residential EV charging in Germany | No price signal; only uncontrolled charging is simulated; household peaks rise by up to a factor of 3.6 |
| Kaschub et al. (2016) | PV-battery households with and without EVs in Germany, under a two-block TOU rate | No local grid metric; the peak is reported per household; it rises by at most 5 % |
| Metcalfe et al. (2026) | Field experiment with an AI-managed EV charging tariff following wholesale prices, 13,233 participants in Great Britain | No local grid metric; the study reports household demand only; peak-hour demand falls by 42 % |
| Valogianni et al. (2026) | Capacity-based pricing for smart charging of 1,000 EVs in the Netherlands | The benchmark price reflects residual grid capacity, not a wholesale-derived energy price; the metric is the aggregate peak of the fleet |
| Voß et al. (2026) | Rural Irish pole transformer with six homes and EVs, charging window 2 a.m. to 4 a.m. | The charging window is imposed, not a response to prices; the synchronized peak follows by assumption |
| Fett et al. (2021) | Diffusion of German home batteries, coupled with an agent-based wholesale market model | No time-varying retail price and no grid metric; the batteries run for self-consumption. Wholesale prices respond to the fleet |
| Avau et al. (2021) | Flemish prosumer with PV, battery and heat pump under day-ahead prices and five distribution tariff designs | Single household, no local grid metric |
| Nazir and Hiskens (2017) | Transactive feeder market for 1,000 thermostatically controlled loads | The price is a local clearing price that caps load at the feeder limit by construction; the study reports load synchronization and oscillations, not peaks |
| Pena-Bello et al. (2023) | PV and home battery sizing under alternative tariffs on three Swiss low-voltage networks | Under the time-of-use tariffs almost no batteries are installed; transformer overloads stem from PV export under every tariff |
| Han et al. (2022) | Stylized population of households with home batteries under a synthetic intraday price fitted to German prices | No local grid and no stated population size; the metric is the exceedance probability of aggregate load. Synchronous peaks of about 5 times mean load are likely |
| Hofmann and Lindberg (2024) | Norwegian households on spot price contracts, winter 2021/22, with automated EV charging | No grid metric; the outcome is the load of household subgroups. The night peak of households with automated EV charging stays below the peak of households without it |
| Kahn-Lang et al. (2025) | Meta-analysis of 140 analyses of time-based rates in the US | No grid metric; the outcome is per-participant peak demand at system peak events |
| Bartusch et al. (2011) | Demand-based distribution tariff in Sweden, 500 households | No time-varying energy price. The shifted load forms a new aggregate peak right after the tariff window |
| Morell-Dameto et al. (2024) | Time-varying network charges and a local network capacity market for heat pumps in Slovenia | No time-varying energy price. Time-differentiated network charges synchronize the heat pump response |
| Kirchem et al. (2025) | Solar prosumage under real-time pricing in a German transmission grid model with endogenous prices | Transmission level, no distribution grid metric. Prosumer peaks rise under real-time pricing |
| Arnold et al. (2024) | EV flexibility flattening national or regional net load in 401 German districts | No price signal and no distribution grid metric; the outcome is net load per district. National incentives raise regional peaks |
| Neetzow et al. (2019) | PV-battery prosumers facing nodal wholesale prices in a German two-level equilibrium model with 16 state-level distribution grids | No local grid metric; the outcome is the aggregate distribution grid capacity per federal state. Prices are endogenous; price-driven storage discharge raises peak feed-in above the case without storage |
| Hofmann et al. (2025) | Five network tariff designs with simulated load shaving and load shifting for 3,608 customers below one Oslo substation | No time-varying energy price and no batteries or EVs; the signal is the network tariff and the flexibility is generic. Under load shifting, the ex-ante time-of-use and critical peak network tariffs raise the substation peak |
| Dyson (2026) | Smart meter data of Queensland households with and without spot price exposure on one day with an evening price spike | No local grid metric; the outcome is the hourly import and export of household groups. Spot-exposed households with large batteries export through the price spike; batteries of customers of major retailers do not |
| Schimeczek et al. (2026), Karaduman (2023), Veenstra and Mulder (2025), Emmanuel and Denholm (2022) | Storage dispatch with price impact in wholesale market models: German day-ahead market, South Australia, Dutch day-ahead and intraday markets, PJM and CAISO | No grid metric, system level. All model the feedback of storage dispatch on wholesale prices |
| Emelianova and Namockel (2025), Guéret et al. (2025) | Flexible EVs and heat pumps, smart charging and vehicle-to-grid in power sector models for Germany and Europe | No grid metric, system level; the outcomes are welfare redistribution, system cost and prices |
| Günther et al. (2021), Lu et al. (2024) | Prosumage under real-time pricing in equilibrium models of the German power sector and of Belgium with 4.8 million households | No local grid metric; the outcome is the net load of a representative household (Günther et al.) or the aggregate residential peak (Lu et al.). Prices are endogenous; real-time pricing does not lower peak withdrawal (Günther et al.). The granularity of retail prices has no systematic effect on peaks (Lu et al.) |
| Gorrasi et al. (2024) | Market designs for low-voltage networks with endogenous residential PV and battery investment under real-time prices: static limits, dynamic operating envelopes, distribution locational prices | Network limits are enforced in every design, so new peaks cannot occur by construction. Prices are endogenous |
| Aniello and Bertsch (2023), Babacan et al. (2017), Forrester et al. (2022) | PV-battery households under dynamic or time-of-use rates in Germany, Australia and the US | Single households or per-customer metrics, no local grid metric. Under time-of-use, arbitrage raises the peak net demand of most customers (Babacan et al.); unconstrained grid discharging raises maximum hourly exports (Forrester et al.) |
| Gebbran et al. (2021), Seyf Abad et al. (2022) | PV-battery prosumers under time-of-use rates on Australian feeders and on 56 UK low-voltage feeders | The battery effect on voltage is not separated from PV (Gebbran et al.); the time-of-use tariff is a network tariff and gives the same PV hosting capacity as a flat rate (Seyf Abad et al.) |
| Backe et al. (2020) | Two campus prosumers with batteries and EV charging under spot prices and capacity subscription network tariffs in Norway | Non-residential, two customers; no spot-only case with a peak result |
| Brinkel et al. (2024) | Smart charging of public on-street EV chargers in a residential low-voltage grid in Utrecht | Grid results only for peak-minimizing charging; the day-ahead price case reports charging cost only |
| Zurmühlen et al. (2026) | Home, feeder and district batteries in German low-voltage feeders | Home batteries run for self-consumption; the only price-driven unit is a district battery with a transformer guard, so no new peak can occur by design |
| Cutter et al. (2021) | Distribution grid cost of transportation electrification in the US | Time-of-use charging is not simulated; the off-peak shift is an assumed 80 % of the coincident EV peak |
| Rapson and Shaffer (2025) | Reanalysis of the Alberta field experiment with EV owners | Same data as Bailey et al. (2025), no independent evidence. Under time-of-use, the maximum transformer load is about 25 % higher than under flat pricing |
| Zadoks et al. (2026) | Time-of-use and congestion-based real-time network tariffs for heat pumps, EVs, PV and batteries in 471 Swiss low-voltage networks | Time-varying network tariffs set by the distribution utility, not wholesale-based retail prices; the real-time price follows forecast transformer loading |
| Samm et al. (2025) | Spot-based import and feed-in tariffs and a time-of-use network charge for one Austrian household with PV and home battery | Single household; the grid metric is the household's own import and feed-in peak. Spot-based rates raise the annual import maximum from 5 kW without battery to 12 kW |
| Verhoeven et al. (2026) | Four static network tariff designs for households with EVs, heat pumps and PV on twelve Dutch low-voltage grids, with fixed or day-ahead energy contracts | No home batteries and no baseline without price response; the study compares tariff designs against a case without tariff instrument under the same day-ahead price exposure. Bandwidth and time-of-use network tariffs raise cable overloading by up to about 170 % through synchronization with low prices |
| Schmeing et al. (2025) | Time-variant against reduced flat network charges under §14a EnWG for heat pumps, batteries and EVs on a German rural low-voltage grid | No baseline without flexibility; the study compares two network charge modules. The time-variant fee cuts congested time steps by 15 % but creates secondary peaks outside the high-tariff windows |
| Zhou et al. (2026) | Closed-loop day-ahead price signal for home energy management systems with HVAC, flexible loads, PV and batteries on a US distribution feeder | The signal is a utility price designed to smooth substation demand and updated from the previous day's demand, not a retail price derived from the wholesale market |
| Strobel et al. (2022) | National against county-level optimization of 10 million EVs in Germany in 2030 | No price signal and no distribution grid: charging minimizes squared net load as a planner; the metric is peak net load at transmission substations. National smoothing, which the authors read as an extreme case of price-sensitive charging, raises average substation peaks by 23 %, more than uncontrolled charging (19 %) |
| Gschwendtner et al. (2025) | Spatially differentiated time-of-use tariffs with and without a power-based capacity price for EV charging at home, work and public locations in the Canton of Zurich | No grid metric; the study reports charging load profiles of the EV population in figures only. Without a capacity price all reaction scenarios form a rebound peak at the start of the low-price period |
| Gomes et al. (2024) | EV-only time-of-use rates via submetering, V2G and network tariff design in a bilevel model with four representative agents | No local grid metric; the outcome is the network cost of four agents; energy prices are fixed time-of-use profiles |
| Li and Jenn (2024) | Feeder overloads from uncontrolled EV charging on more than 5,000 feeders of the three Californian investor-owned utilities, 2025 to 2045 | No price signal; charging sessions are bootstrapped from empirical records. Overloads on home-charging feeders cluster after 6 p.m. and are most likely around midnight |
| Qiu et al. (2022) | Smart meter data of about 1,600 EV homes and 2,224 non-EV homes in Arizona, 2014 to 2019, on block, time-of-use and EV time-of-use rates | No local grid metric; the outcome is household load by hour. Customers on the EV time-of-use rate add load mostly from 11 p.m. to 5 a.m. |
| Li et al. (2026) | Hourly charging of about 20,900 private residential EV chargers in Shandong, China, before and after voluntary time-of-use adoption, 2023 | No grid metric; the outcome is charging volume by tariff period. Off-peak charging rises and peak-period charging falls, concentrated from 11 p.m. to 5 a.m. |
| Brucke et al. (2026) | Time-varying network charges derived from line utilization for a standalone storage unit of 125 MW and 250 MWh at one 110 kV node, three German regions, 2045 scenario | Large-scale storage at 110 kV, no low-voltage grid metric; the metric is the required line capacity in MW km. The price is the normalized net load, exogenous because the storage is assumed small relative to the system. Price-only operation raises grid needs in all three regions; the daily repeating §14a network charges raise them in the wind region |

# A3 Flexibility contracts

Table A3. Examples of flexibility contracts (as of September 2026)

| Offer | Market, status | Asset | Customer receives | Customer commits to |
|---|---|---|---|---|
| Octopus Power Pack | UK; tariff, beta since 2024 | EV (BYD Dolphin) with V2G charger | Free home charging up to 210 kWh per month | Plugged in 12 h per day on 20 days per month; minimum state of charge below 30 % |
| Octopus and BYD, Power Pack Bundle | UK; car and tariff bundle, 2025 | Leased EV, V2G charger | Car lease below £300 per month including free home charging | Plugged in overnight |
| Octopus Powerloop | UK; trial 2020–2022, 135 customers | Leased EV (Nissan Leaf), V2G charger | Car and bidirectional charger | Not specified in source |
| Renault, Mobilize and The Mobility House, Mobilize Power | France; since 2024 | EV (Renault 5, Alpine A290), bidirectional AC charger | Free charging | Mobilize charger and supply contract |
| The Mobility House, V2G tariff | Germany; announced 2026 | EV (Renault 5; later Mercedes GLC, CLA) | Free charging electricity | Plugged in every night; minimum state of charge set in advance (e.g., 60 % at 7:00) |
| BMW and E.ON, V2G tariff | Germany; since 2026 | EV (BMW iX3), bidirectional wallbox | 0.24 EUR per plugged-in hour, up to 60 EUR per month | Supply and feed-in contract; smart meter |
| Mercedes-Benz, MB.CHARGE Home | Germany, France, UK; planned 2026 | EV (Mercedes GLC), bidirectional charger | Not published | Minimum range and departure time set by user |
| 1KOMMA5°, Heartbeat | Sweden; since 2023 | Home battery with PV | Electricity at 0 EUR/kWh up to 15,000 kWh per year | Battery pooled by provider; monthly software fee |
| Octopus Zero Bills | UK; new-build homes | PV, home battery, heat pump | No energy bill for a guaranteed period, fair-use terms | Provider operates all assets |
| Octopus, Intelligent Octopus (battery) | Germany | Home battery | Bonus of 0.10 EUR/kWh on grid-charged energy | Provider controls charging only; smart meter |
| Green Mountain Power, energy storage lease | Vermont, US | Home battery (Tesla Powerwall) | Battery lease at 55 USD per month over ten years; backup power | Utility discharges the battery at peak times |
| SOLRITE and sonnen | Texas, US; 2025 | PV and two home batteries | Hardware at no upfront cost; solar electricity at 0.12 USD/kWh | Provider operates batteries in ERCOT |

# A4 Model description

This section states the optimization problems of households and utility-scale storage and the equilibrium condition. Table A4 defines the symbols. All problems are solved for each calendar month separately.

Each household $i$ with a battery minimizes its electricity bill:

$$\min \sum_t \left[(p_t + \tau) w_{i,t} - p_t e_{i,t} + (p_t + \tau)\left(c_{i,t}^r - d_{i,t}^r\right) + \kappa\left(d_{i,t} + d_{i,t}^r\right)\right] \tag{A1}$$

Withdrawal and injection balance load, PV generation and the own storage account:

$$w_{i,t} - e_{i,t} = l_{i,t} - v_{i,t} + c_{i,t} - d_{i,t} \tag{A2}$$

The own account holds solar or grid electricity for consumption in the household. The refund account holds grid electricity for reinjection:

$$s_{i,t} = s_{i,t-1} + \eta c_{i,t} - d_{i,t}/\eta \tag{A3}$$

$$s_{i,t}^r = s_{i,t-1}^r + \eta c_{i,t}^r - d_{i,t}^r/\eta \tag{A4}$$

Both accounts share the power and energy limits of the battery. Within an hour, charging and discharging share the available time:

$$c_{i,t} + d_{i,t} + c_{i,t}^r + d_{i,t}^r \leq \bar{P} \tag{A5}$$

$$s_{i,t} + s_{i,t}^r \leq \bar{E} \tag{A6}$$

All variables are non-negative. Storage is empty at the start of each month:

$$s_{i,0} = s_{i,0}^r = 0 \tag{A7}$$

Home battery dispatch in Eq. (3) is the sum over both accounts:

$$d_{i,t}^h = d_{i,t} + d_{i,t}^r, \;\; c_{i,t}^h = c_{i,t} + c_{i,t}^r \tag{A8}$$

Utility-scale storage maximizes its arbitrage revenue:

$$\max \sum_t [p_t(d_t^u - c_t^u) - \kappa^u(c_t^u + d_t^u)] \tag{A9}$$

subject to storage dynamics and limits:

$$s_t^u = s_{t-1}^u + \eta c_t^u - d_t^u/\eta, \;\; c_t^u + d_t^u \leq \bar{P}^u, \;\; s_t^u \leq \bar{E}^u \tag{A10}$$

Under exogenous prices, $p_t$ is a fixed time series. Under endogenous prices, $p_t$ follows Eqs. (1) to (4) and all agents take it as given. The equilibrium is computed as one linear program per month that minimizes

$$\sum_t \int_0^{R_t} f(x)\mathrm{dx} + \sum_t \varepsilon_t R_t + N \sum_i \sum_t \left[\tau w_{i,t} + \tau\left(c_{i,t}^r - d_{i,t}^r\right) + \kappa\left(d_{i,t} + d_{i,t}^r\right)\right] + \sum_t \kappa^u (c_t^u + d_t^u) \tag{A11}$$

subject to Eqs. (2), (3) and (A2) to (A10). Its optimality conditions coincide with those of problems (A1) and (A9) at $p_t = f(R_t) + \varepsilon_t$. The integral is approximated piecewise linearly with 256 segments.

For the test without coordination (Section 6.3), the range of cost-minimizing dispatch of each household in each hour is determined from 40 cost-minimizing schedules per month. These are obtained by solving problem (A1) at the equilibrium price with a small random perturbation of the price in each hour and of the cost of charging. In each hour, each household dispatches at the lower or the upper end of its range. The probability of the upper end is the same for all households and is set such that the expected fleet dispatch equals the equilibrium dispatch. Draws are independent across households and hours; 1,000 networks are simulated.

Table A4. Symbols

| Symbol | Definition | Unit or value |
|---|---|---|
| $t$ | Hour of the month | h |
| $i$ | Household with a battery | – |
| $H$ | Number of households with a battery per grid | 0 to 50 |
| $p_t$ | Wholesale price | EUR/MWh |
| $\tau$ | Network charges and taxes on withdrawal | 0.10 EUR/kWh |
| $\kappa, \kappa^u$ | Cycling cost of home batteries, per unit discharged, and of utility-scale storage, per direction | 20 EUR/MWh each |
| $\eta$ | One-way efficiency | $\sqrt{0.9}$ |
| $\overline{P}, \overline{E}$ | Power and capacity of a home battery | 4 kW, 8 kWh |
| $\overline{P}^u, \overline{E}^u$ | Power and capacity of utility-scale storage | 10 GW, 60 GWh |
| $l_{i,t}, v_{i,t}$ | Household load and PV generation | kW |
| $w_{i,t}, e_{i,t}$ | Withdrawal and injection, excluding the refund account | kW |
| $c_{i,t}, d_{i,t}$ | Charging and discharging, own account | kW |
| $c^r_{i,t}, d^r_{i,t}$ | Charging and discharging, refund account | kW |
| $c^h_{i,t}, d^h_{i,t}$ | Charging and discharging of a home battery, both accounts (A8) | kW |
| $s_{i,t}, s^r_{i,t}$ | State of charge, own and refund account | kWh |
| $c^u_t, d^u_t, s^u_t$ | Charging, discharging and state of charge of utility-scale storage | GW, GWh |
| $L_t, W_t, S_t$ | National load (including pumped-storage losses and the daily mean of net exports), wind and solar generation | GW |
| $b_t$ | Battery dispatch: discharging minus charging of utility-scale storage and of the home batteries in all N grids | GW |
| $R_t$ | Net load after flexibility (including home storage) | GW |
| $f$ | Price function | EUR/MWh |
| $a$ | Parameter of the cubic price function (4) | $1.8 \times 10^{-3}$ EUR/MWh per $GW^3$ |
| $a_2$ | Parameter of the quadratic price function (5) | $8.9 \times 10^{-2}$ EUR/MWh per $GW^2$ |
| $\varepsilon_t$ | Price residual: observed price minus price function at thermal net load, fixed across adoption levels | EUR/MWh |
| $N$ | Number of identical local distribution grids | 190,000 |

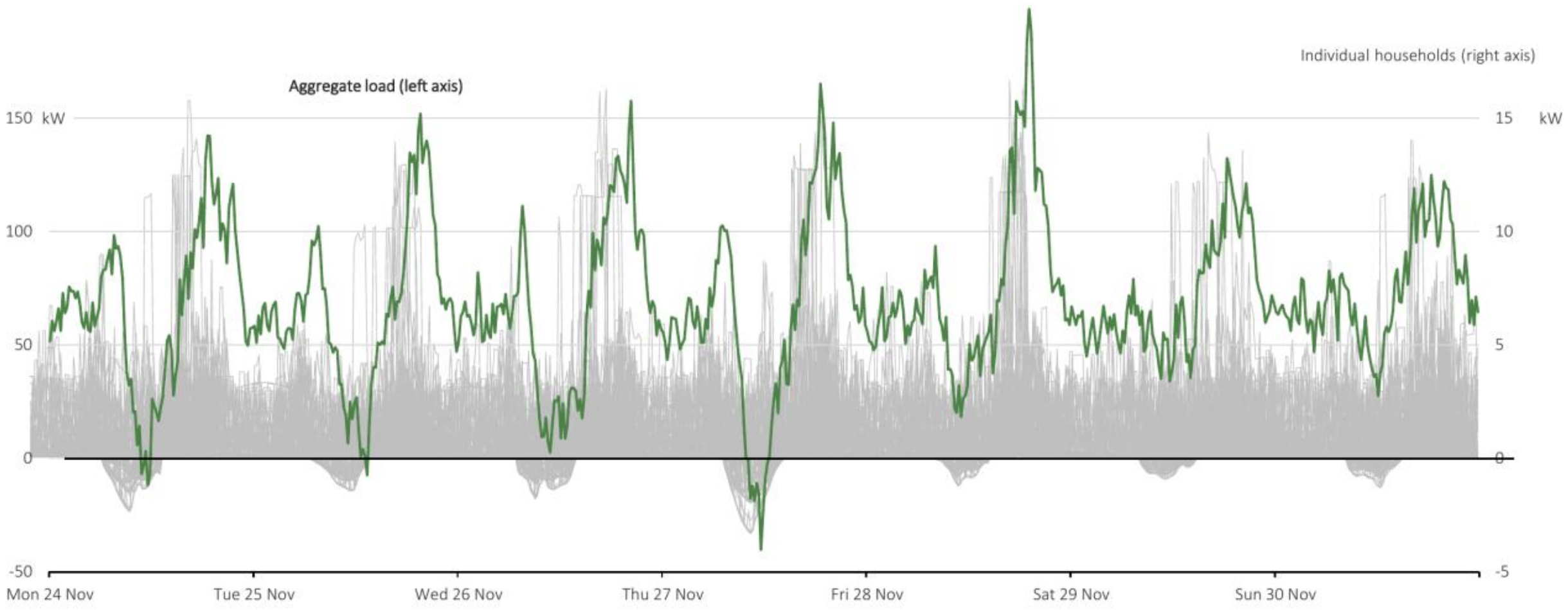

Figure A1. Individual household load and aggregate load during the week of the annual import peak.

# A5 Additional results

Table A5. Coincident peak at full adoption under exogenous and endogenous prices, all model runs (change relative to no-battery peak)

| Sensitivity | Description | Import peak | | Export peak | |
|---|---|---|---|---|---|
| | | Exo-genous prices | Endo-genous prices | Exo-genous prices | Endo-genous prices |
| Base case | 50 % of households with a home battery; mean load in the highest 1 % of hours | +83 % | -6 % | +1 % | -15 % |
| **Battery size and adoption** (base case: 8 kWh / 4 kW; cycling cost 20 EUR/MWh on discharge) | | | | | |
| Larger batteries | Battery size doubled (16 kWh / 8 kW) | +224 % | +12 % | +105 % | +22 % |
| Smaller batteries | Battery size halved (4 kWh / 2 kW) | +14 % | -9 % | -8 % | -10 % |
| High power | Power doubled, energy unchanged (8 kWh / 8 kW) | +214 % | -5 % | +82 % | -7 % |
| Low power | Power halved, energy unchanged (8 kWh / 2 kW) | +17 % | -7 % | -22 % | -23 % |
| High cycling costs | Cycling cost of 100 EUR/MWh on discharge (home batteries) and 50 EUR/MWh per direction (utility-scale storage) | +26 % | -5 % | -11 % | -14 % |
| All households | All households adopt batteries, including those without solar PV | +224 % | +15 % | +105 % | +23 % |
| **Local concentration and utility-scale flexibility** (base case: $N = 190{,}000$; utility-scale flex 10 GW / 60 GWh) | | | | | |
| Quarter of the grids | National home battery fleet 25 % of base case | +83 % | +24 % | +1 % | -9 % |
| Half of the grids | National home battery fleet 50 % of base case | +83 % | +4 % | +1 % | -10 % |
| Larger national fleet | National home battery fleet 200 % of base case | +83 % | -8 % | +1 % | -24 % |
| No utility-scale flex | No utility-scale storage | +83 % | -3 % | +1 % | -14 % |
| Less utility-scale flex | Utility-scale flexibility halved to 5 GW / 30 GWh | +83 % | -5 % | +1 % | -13 % |
| More utility-scale flex | Utility-scale flexibility doubled to 20 GW / 120 GWh | +83 % | -7 % | +1 % | -17 % |
| **Renewables** (base case: 8 kWp per household; no curtailment) | | | | | |
| PV curtailment | Economic curtailment of residential PV | +84 % | -4 % | 0 % | -18 % |
| Larger residential PV | PV size doubled to 16 kWp per solar household | +81 % | -10 % | -7 % | -12 % |
| More solar | Additional 50 TWh/a of solar PV (non-residential) | +83 % | -5 % | +1 % | -19 % |
| More wind | Additional 50 TWh/a of onshore wind | +83 % | +4 % | +1 % | -13 % |
| Future power system | 2035 scenario: wind 200 GW, solar 250 GW, load 750 TWh, utility-scale flexibility 33 GW / 200 GWh | +90 % | +8 % | +3 % | -16 % |
| **Network charges** (base case: network charges and taxes 0.10 EUR/kWh; refund on reinjected electricity) | | | | | |
| Low network charge | Network charges and taxes of 0.05 EUR/kWh | +86 % | -4 % | +4 % | -14 % |
| High network charge | Network charges and taxes of 0.20 EUR/kWh | +78 % | -7 % | -6 % | -16 % |
| No charge refund | No refund of the network charge on reinjected electricity | +30 % | -9 % | -12 % | -16 % |
| Net metering | Network charge of 0.10 EUR/kWh only on each household's monthly net withdrawal (withdrawal minus injection); no refund | +88 % | +3 % | +8 % | -8 % |
| Injection charge | Network charge of 0.10 EUR/kWh also on injection; refund on reinjected electricity unchanged | +82 % | -7 % | 0 % | -15 % |
| **Price function** (base case: exogenous: historical 2025 prices; endogenous: cubic price function) | | | | | |
| Cubic | Exogenous: cubic price function at $X = 0$; endogenous: cubic | +85 % | -6 % | +1 % | -15 % |
| Quadratic | Exogenous: quadratic price function at $X = 0$; endogenous: quadratic | +86 % | -7 % | +2 % | -20 % |
| Piecewise linear | Exogenous: piecewise linear price function at $X = 0$; endogenous: piecewise linear | +85 % | -7 % | 0 % | -25 % |
| Cubic, ε = 0 | Exogenous: same function at zero batteries; endogenous: cubic, price residual $\varepsilon$ set to zero | +85 % | -11 % | -10 % | -28 % |
| Flatter price function | Exogenous: same function at zero batteries; endogenous: cubic, parameter $a$ halved | +84 % | -3 % | -1 % | -12 % |
| Steeper price function | Exogenous: same function at zero batteries; endogenous: cubic, parameter $a$ doubled | +89 % | -7 % | +3 % | -17 % |
| **Definition of the coincident peak** (base case: mean load in the highest 1 % of hours) | | | | | |
| Peak hour | Load in the highest hour of the year | +65 % | +3 % | +12 % | -8 % |
| Top 10 % | Mean load in the highest 10 % of hours | +42 % | -7 % | -3 % | -26 % |

| Household simulation (base case: 100 households, single draw, perfect foresight within each month, hourly) | | | | | |
|---|---|---|---|---|---|
| Bootstrap | 400 replications, stratified by solar ownership (50 solar households and 50 households without solar), from the base dispatch; mean ± standard deviation | +74 ± 11 % | -6 ± 2 % | +1 ± 3 % | -14 ± 2 % |
| Independent networks | Three new networks (new load profiles and coupling), separate equilibrium for each; range | +81 to +95 % | -8 to -5 % | 0 % | -15 to -14 % |
| Network size | 25 to 400 households, national fleet unchanged; range | +65 to +94 % | -7 to -5 % | 0 to +1 % | -15 to -13 % |
| No load coupling | Household load not coupled to national load | +81 % | -6 % | +1 % | -14 % |
| Stronger load coupling | Coupling of household load to national load at double strength | +77 % | -6 % | 0 % | -15 % |
| Measured load Zurich | Measured household load (EKZ Zurich, 2024, annual energy normalized) instead of synthetic profiles; against synthetic profiles for 2024 | +133 % | +23 % | +20 % | -10 % |
| Measured load Lucerne | Measured household load (CKW Lucerne, 2025, annual energy normalized) instead of synthetic profiles | +116 % | +6 % | +14 % | -13 % |
| Daily foresight | Home batteries with a one-day horizon (empty at midnight), utility-scale flexibility unchanged | +87 % | -1 % | +3 % | -15 % |
| Uncoordinated households | Households choose at random within their range of cost-minimizing dispatch; mean of 1,000 networks | n.a. | -4 % | n.a. | -14 % |
| Quarter-hourly | All quantities quarter-hourly; the year solved as twelve monthly problems | +77 % | -7 % | +1 % | -14 % |
| Historical year | Load, solar, wind and prices of 2023 or 2024 instead of 2025; price function estimated for each year; range | +76 to +79 % | -3 to -1 % | +9 to +11 % | -12 to -11 % |
| PV orientation | All PV facing south, or half east and half west (base case: east, south, west and flat); range | +83 to +85 % | -8 to -5 % | -6 to +20 % | -14 to -9 % |
| Fewer solar households | 30 % of households with PV (base case: 50 %) | +28 % | -8 % | -4 % | -14 % |

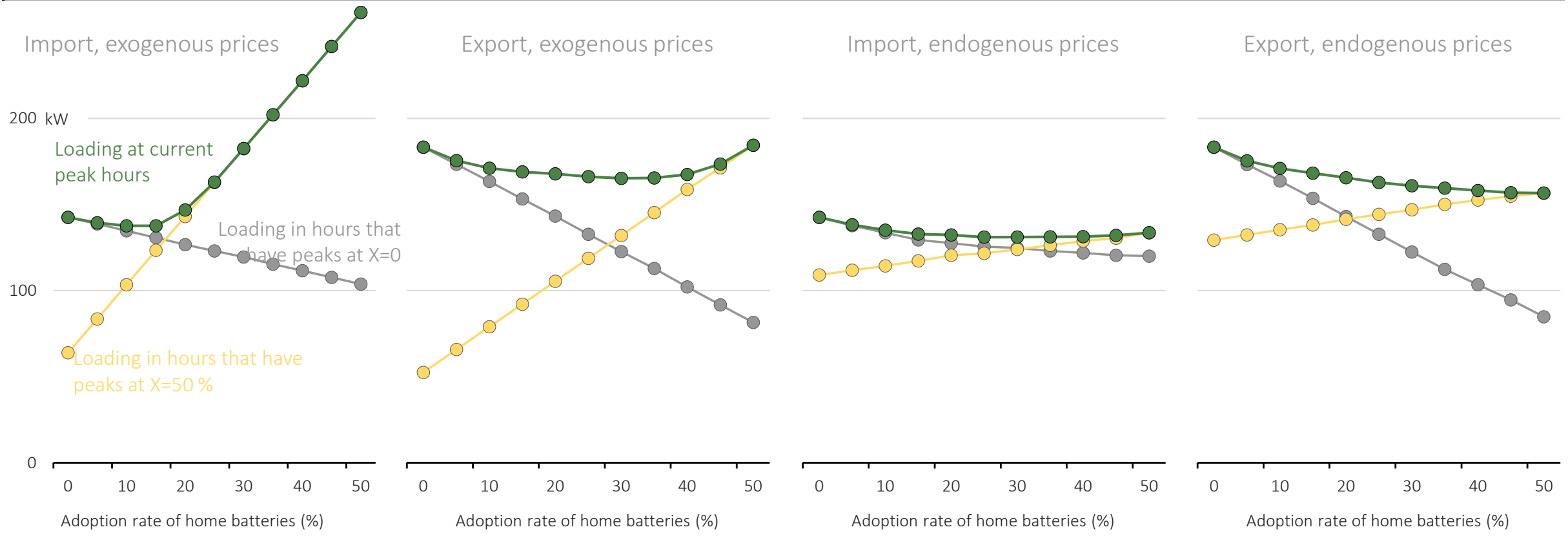

Figure A2. Loading in the current peak hours, in the peak hours without batteries and in the peak hours at full adoption (top 1 % of hours).

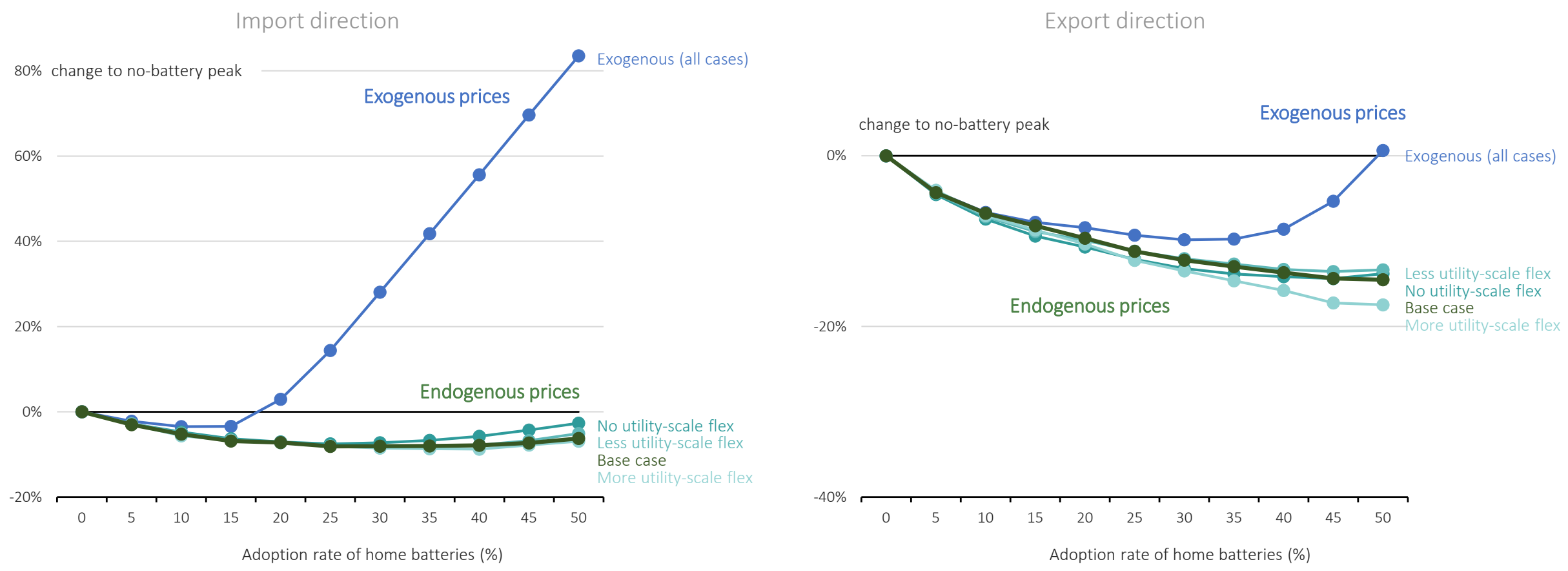

Figure A3. Coincident peak for exogenous and endogenous prices: utility-scale flexibility.

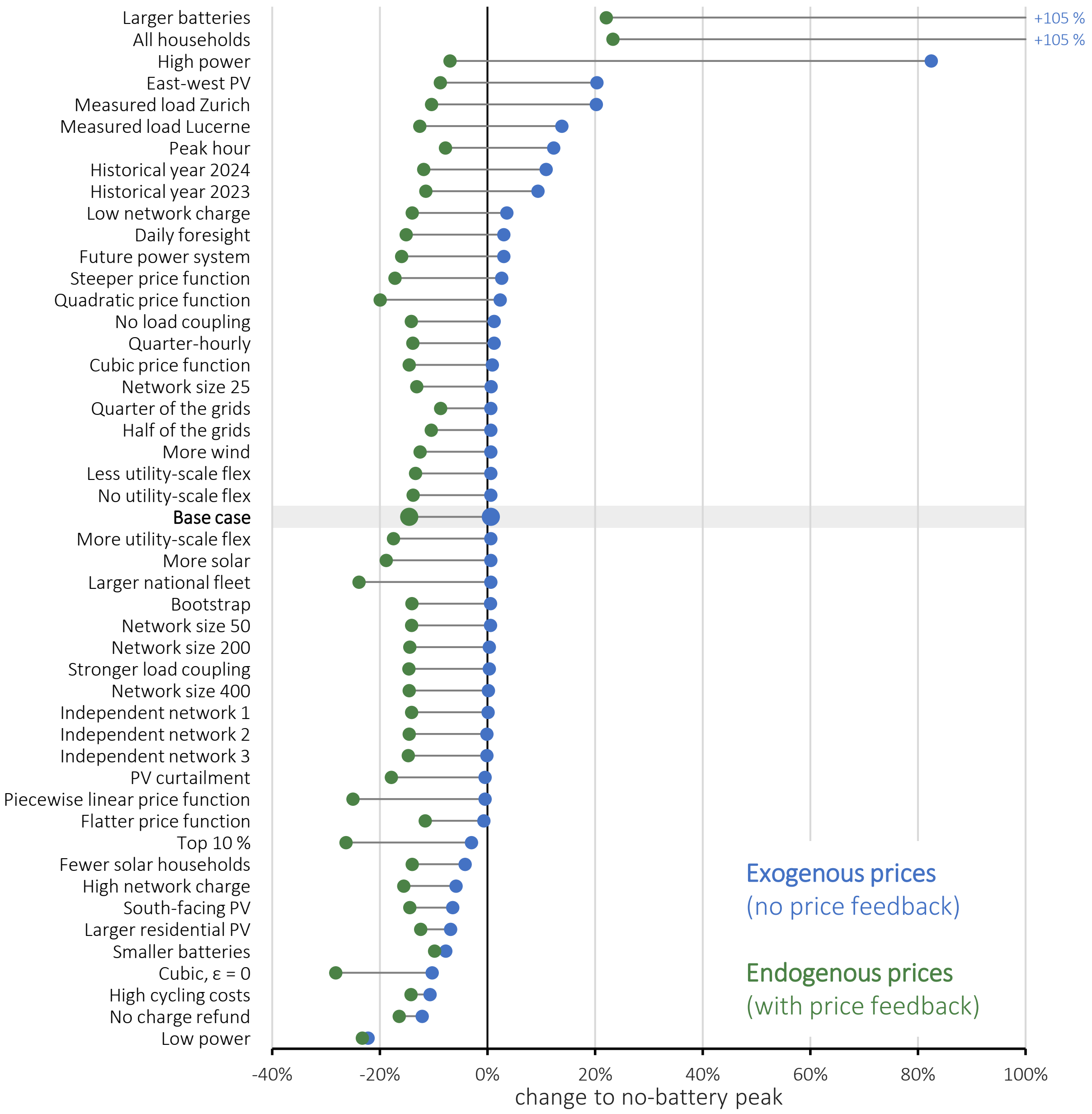

Figure A4. Coincident peak at full adoption across sensitivities, export direction (import direction: Figure 11).

# A6 References cited only in the appendix

# Statements

## Acknowledgments

I thank Johanna Bronisch, Anselm Eicke and Clemens Lohr for helpful discussions. I thank Sabine Anatolitis-Pelka, Christian Bantle, Joachim Benatzky, Karoline Brucke, Simon Fischer, Reyer Gerlagh, John Henningsen, David Holmberg, Lena Hörnlein, Pei Huang, Magnus Jennerholm, Patrick Jochem, Jan Horst Keppler, Lukas Küng, Ioannis Lampropoulos, Marcel Linnemann, Robert Metcalfe, Helene Neugebauer, Carlo Obersteiner, Marten Ovaere, Danny Pudjianto, Erik Richter-Alten, Oliver Ruhnau, Alexander Sauer, Jakob Schlandt, Wouter Schram, David Schröder, Simon Tindemans, Floris van Montfoort, Gregor Verbič, Alexander Vogt, Thomas Walter, Yan Wu and Jan Zacharias for helpful comments.

## Competing interest

The author is managing director of Neon Neue Energieökonomik, a consultancy advising clients in the electricity sector. This paper did not receive any funding and was not commissioned by any client; no client has seen its results prior to publication.

## Funding

This research did not receive any specific grant from funding agencies in the public, commercial, or not-for-profit sectors.

## Data availability

The simulation code, all input data and the complete simulation results are available from the author on request and will be deposited in a public repository upon acceptance.

## CRediT contribution statement

Lion Hirth: Conceptualization, Methodology, Software, Formal analysis, Investigation, Writing – original draft, Writing – review & editing, Visualization.

## AI

During the preparation of this work the author used Claude (Anthropic) to draft and edit text, write simulation and analysis code, and produce tables and figures. After using this tool, the author reviewed and edited the content as needed and takes full responsibility for the content of the published article.